\documentclass{aa}  

\usepackage{graphicx}
\usepackage{txfonts}
\usepackage[]{hyperref}
\usepackage{epstopdf}
\usepackage{todonotes}
\usepackage{multirow}
\usepackage{xcolor}
\begin{document}

   \title{I. Analysis of candidates for interacting galaxy clusters}

   \subtitle{A1204 and A2029/A2033}

  \author{Elizabeth Johana Gonzalez \thanks{elizabethjgonzalez@oac.unc.edu.ar }\inst{1,2} \and Mart\'in de los Rios\inst{1,2} \and Gabriel A. Oio\inst{1,2} \and Daniel Hern\'andez Lang\inst{4} \and
   Tania Aguirre Tagliaferro\inst{1,2} \and Mariano J. Dom\'inguez R.\inst{1,2} \and Jos\'e Luis Nilo Castell\'on\inst{3,4} \and H\'ector Cuevas L.\inst{4} \and Carlos A. Valotto\inst{1,2}}

\institute{Instituto de Astronom\'{\i}a Te\'orica y Experimental, (IATE-CONICET),
 Laprida 854, X5000BGR, C\'ordoba, Argentina.
         \and
   Observatorio Astron\'omico de C\'ordoba, Universidad Nacional de C\'ordoba, Laprida 854, X5000BGR, C\'ordoba, Argentina.
   \and
    Instituto de Investigación Multidisciplinario en Ciencia y Tecnología, Universidad de La Serena. Benavente 980, La Serena, Chile. 
   \and  
    Departamento de F\'isica y Astronom\'ia, Facultad de Ciencias, Universidad de La Serena. Av. Juan Cisternas 1200, La Serena, Chile.
             }
             
   \date{2017}

\abstract{Merging galaxy clusters allow for the study of different mass components, dark and baryonic, separately. Also, their occurrence enables to test the $\Lambda$CDM scenario, which can be used to put constraints on the self-interacting cross-section of the dark-matter particle.}{It is necessary to perform a homogeneous analysis of these systems. Hence, based on a recently presented sample of candidates for interacting galaxy clusters, we present the analysis of two of these cataloged systems.}{In this work, the first of a series devoted to characterizing galaxy clusters in merger processes,  we perform a weak lensing analysis of clusters A1204 and A2029/2033 to derive the total masses of each identified interacting structure together with a dynamical study based on a two-body model. We also describe the gas and the mass distributions in the field through a lensing and an X-ray analysis. This is the first of a series of works which will analyze these type of system in order to characterize them.}{Neither merging cluster candidate shows evidence of having had a recent merger event. Nevertheless, there is dynamical evidence that these systems could be interacting or could interact in the future.}{It is necessary to include more constraints in order to improve the methodology of classifying merging galaxy clusters. Characterization of these clusters is important in order to properly understand the nature of these systems and their connection with dynamical studies. } 
 

   \maketitle
%

%

\section{Introduction}

Matter components in galaxies and galaxy systems contribute to the common gravitational potential with roughly concentric distributions. This matter distribution hinder the analysis of dark- and baryonic-mass distributions independently. Interacting galaxy clusters allow us to study the different mass components separately, given that, in the standard ``cold dark matter'' model, the intracluster gas and the dark matter component can be spatially segregated in the merging process \citep{Furlanetto2002}. The galaxies and the dark matter of the merging systems behave as collisionless particles, while the gas component is fluid-like and experiences ram pressure. Therefore, the analysis of merging clusters provides independent information regarding the baryonic component, mostly constituted by the intracluster plasma, and the gravitational potential traced by the dark-matter distribution.\\
Merging galaxy clusters are the largest and most energetic events in the Universe since the Big Bang \citep{Sarazin2002}. Also, they provide evidence that supports the existence of dark matter \citep{Dawson2013, Dahle2013, Harvey2015} since, when clusters collide, galaxies behave as collisionless particles while the X-ray emission plasma is slowed by ram pressure. This scenario allows us to test the existence of dark matter. Without dark matter, the gravitational potential would be traced by the dominant baryonic component, the X-ray plasma. On the other hand, if the principal component in clusters was collisionless dark matter then the gravitational potential would be traced by this component which is expected to follow the galaxy distribution. Thus, deriving the gravitational potential distribution allows us to distinguish between these possibilities. \\
 One of the hottest and most X-ray-luminous
clusters observed is 1E0657-558, the \textit{Bullet cluster}. This cluster, at $z=0.296$, was discovered by \citet{Tucker1995} using \textit{Chandra} X-ray observations. The most probable scenario would be that the Bullet cluster is the result of the merging process of two systems, taking place in the sky plane \citep{Barrena2002,Markevitch2002}. X-ray observations revealed the presence of a fairly cold ($T \sim 6\,KeV$) bullet-like structure, just leaving the core of the dominant system ($T \sim 14\,KeV$) with a velocity of $4500\,km\,s^{-1}$, producing a prominent bow shock \citep{Markevitch2002}. Taking into account that the offset between both systems is $0.66\,Mpc$, the closest approach would have occurred $0.1-0.2$\,Gigayears ago. Weak lensing analysis of this system showed that the gravitational potential does not follow the plasma distribution, but rather is consistent with the galaxy distribution \citep{Clowe2006}. The discrepancy between the total mass center and the baryonic center provides evidence for the existence of dark matter with an $8\sigma$ confidence level.\\
Several works have analyzed Bullet-like clusters obtaining direct empirical proof of the existence of dark matter \citep{Jee2014,Bradac2008,Dawson2012,Andrade2015,vanWeeren2017}. Even though the standard cosmological model has proved to be successful at large cosmological scales, there have been differences between the predictions of this model and the observations at smaller scales. These disagreements could be overcome if dark-matter particles were self-interacting \citep{Spergel2000,Rocha2013,Zavala2013}. The analysis of merging systems could provide an upper limit for the ratio between the self-interacting cross-section and the mass of the dark-matter particle, which could be obtained by taking into account the separations of the matter components after the interaction between the galaxy systems. \citet{Harvey2015} obtained $\sigma_{DM}/m = -0.25^{+0.42}_{-0.43}$\,cm$^2$/g with a confidence level of 68$\%$, or $\sigma_{DM}/m < 0.47$\,cm$^2$/g with a confidence level of 95$\%$ based on the analysis of 30 merging clusters. This result rules out parts of model space of hidden sector dark-matter models that predicts $\sigma_{DM} \sim 0.6 cm^2/g$ on cluster scales through a long-range force. \\
With the aim of testing the standard cosmological model $\Lambda$CDM, different studies used numerical simulations to  analyze the probability of observing a system with similar properties to those observed in the Bullet Cluster. There is extensive debate regarding this issue. Several works conclude that the velocity observed for the less massive system would not be physically possible in a $\Lambda$CDM scenario \citep{Markevitch2002,Springel2007}. Even taking into account that the matter velocity \citep[$\sim 2600$ km\,s$^{-1}$,][]{Koda2007} would be lower than the plasma velocity, $\Lambda$CDM models would not be able to produce Bullet-like systems \citep{Lee2010,Thompson2012}. Nevertheless, results based on larger cosmological simulations demonstrate that these systems are not extreme, thus, there would be no tension between the observations and the standard cosmological model. Interestingly, \citet{Thompson2012} conclude that a volume of $(4.5\,h^{-1}\,Gpc)^{3}$ would be required in order to observe Bullet-like clusters.\\
Taking into account the importance of the analysis of merging systems, it is necessary to study uniformly selected samples to derive robust constraints. \citet{delosRios2016} have recently presented a method to identify galaxy merging systems based on redshift and photometric galaxy catalogs. This  methodology  provides  a  highly  reliable  sample  of candidates for merging systems with low contamination and some estimated properties. This paper is the first in a series of works that analyzes merging systems cataloged by \citet{delosRios2016}. The analysis includes photometric, lensing, X-ray, radio, Sunyaev-Zeldovich, and dynamical studies. Our aim is to characterize the cataloged merging candidate clusters. \\
In this paper, the first of a series devoted to characterizing galaxy clusters in merger processes, we present a dynamical and a lensing analysis of the galaxy clusters A1204 and A2029/A2033. We derive the lensing masses for each identified interacting structure, then we characterize the dynamical state of these clusters according to a two-body model. We also perform a qualitative description of the gas and the mass distribution in the field of the systems based on a lensing and an X-ray analysis. These systems were randomly selected from the merging system catalog, taking into account the publically available data for the lensing and X-ray analysis. Details regarding the redshift catalogs for the dynamical analysis, the observations used for the lensing, and the X-ray studies are presented in Section \ref{sec:osbservations}. In Section \ref{sec:dynamic} we characterize the dynamic state of these systems. In Section \ref{sec:lensing} we describe the lensing methodology. We present and discuss our results in Section \ref{sec:results} and finally our conclusions are presented in Section \ref{sec:conc}. When necessary, we adopt a standard cosmological model: $H_0 = 70 $km\,s$^{-1}$\,Mpc,$\Omega_m = 0.3$ and $\Omega_\Lambda = 0.7$.

\section{Data acquisition}
\label{sec:osbservations}

\subsection{Photometric observations}

Details of the studied objects and images for the weak lensing analysis are presented in Table \ref{table:1}. There we show the location of the primary components of the merging systems and their corresponding images used for the analysis.\\
For the lensing analysis, in the case of A1204, we use optical images retrieved from the SMOKA (Subaru Mitaka Okayama Kiso Archive) database. Observations were obtained using the Suprime-Cam \citep{Miyazaki2002} with the wide-field prime focus of the 8.2m Subaru telescope located at the Mauna Kea Observatory, under the observation program COSMOS (Cosmic Evolution Survey). The Suprime-Cam is an 80-megapixel ($10240 \times 8192$) mosaic CCD camera, covering $34' \times 27'$ with a resolution of $0.202''/$pixel. Retrieved images are in the i+ and V Subaru filters. Reduction and astrometric calibration is performed using SDFRED2 in a standard mode, by combining 90 and 70 frames for filters $i+$ and $V$, respectively.\\
In the case of A2029/2033, observations for the lensing analysis were retrieved from the Canadian Astronomy Data Centre (CADC), taking into account the optical requirements for the lensing measurements: long exposure times, red optical filters, and a large celestial coverage. Therefore we use reduced and calibrated data observed with the MEGACAM camera mounted on the 3.6m CFHT telescope in $r'$ and $i'$ bands. MEGACAM consists of 36 CCDs with $2048 \times 4612$ pixels with a resolution of $0.187''$, covering a full $1 \times 1$ square degree field of view. Observations were carried out through Queued Service Observing (QSO). The data were pre-processed (removal of the instrumental signature) and calibrated (photometry and astrometry) using Elixir version 2.0. In order to perform the lensing analysis, we use a parallelized pipeline, to carry out the source detection, photometry, classification and the shape measurements in the 36 frames in parallel. 

\begin{table*}
\caption{Galaxy cluster sample and observation specifications.}             
\label{table:1}      
\centering          
\begin{tabular}{c c c c l l l c}     
\hline\hline       
Primary component & $\alpha$ & $\delta$& z & Program & Filters & Exp-time & Seeing \\ 
ID. & J200 & J200 & & ID. & & [seconds] & [arcseconds] \\ 
\hline                    
   A1204 & 11h 13m 32.2s & +17$^\circ$ 35' 40'' & 0.1706  &  o10114 & Subaru $i+$ & 240 & $0.74$\\  
         &               &                      &         &  o11203 & Subaru V & 240 & $0.91$\\  
         &               &                      &         &  WG931430P & ROSAT PSPCC & 14699 & \\       
   A2029 & 15h 10m 58.7s & +05$^\circ$ 45' 42'' & 0.0775  & 06AC16 & CFHT i.MP9701 & 500 & $0.52 \dagger$\\  
         &               &                      &         & 03AC28 & CFHT r.MP9601 & 480 & $0.65 \dagger$\\  
         &               &                      &         & US800249P & ROSAT PSPCB & 12542 & \\  

  \hline                  
\end{tabular}
\medskip
\begin{flushleft}
\textbf{Notes.} Columns: (1) shows the cluster identification; (2), (3) and (4), the coordinates of the center and redshift according to the CDS (Centre de Donn\'ees astronomiques de Strasbourg) database; (5), (6), (7) and (8) image specifications used for the lensing and X-ray analysis. ($\dagger$ Average seeing for the mosaic array).
\end{flushleft}
\end{table*}
\subsection{X-ray observations}

In this work we intend to describe the gas distribution in the field of the analyzed systems from the X-ray brightness contours. Hence, our present work does not focus on the detailed spatial distribution and spectral characteristics of the hot intra-cluster gas. Taking this into account, for our X-ray analysis we choose to make use of the ROSAT data as a compromise between sensitivity, spatial resolution, and field of view.\\ 
Archival  ROSAT \footnote{See \url{http://www.mpe.mpg.de/xray/wave/rosat/index.php}}  imaging  data  for  our  sample  clusters were downloaded from the High Energy Astrophysics Science Archive Research Center (HEASARC) on-line repository. Data acquisition details are showed in Table \ref{table:1}. The ROSAT X-ray data are in the form of event files, which we use to generate images. The ROSAT Position Sensitive Proportional Counter (PSPC) observations in the soft and hard bands were used to construct a broad band image. These data sets were prepared for analysis by first eliminating sources of contamination and selecting good time intervals. We calculate the background, assuming a Poisson statistics with the tool \texttt{ximage} from the HEASARC software, to subtract it from the surface brightness. For the broad band image of A1204 we obtain a root mean square (rms) noise of 0.32 (cts /im pix) and from the data of A2029 the rms noise obtained is of 0.72 (cts / im pix). The X-ray images in the 0.1 – 2.4 keV band are used to measure the X-ray surface brightness profiles of the analyzed clusters.  \\
\subsection{Redshift catalogs}

Here we describe the redshift catalogs used for dynamical analysis. For the A1204 cluster we use the spectroscopic data from the Hectospec Cluster Survey (HeCS). This catalog was constructed by \cite{Rines2013} using the Hectospec instrument \citep{fabricant} mounted on the MMT 6.5m telescope. The HeCS clusters were selected from X-ray catalogs based on the  ROSAT All Sky Survey \citep[RASS, ][]{voges}, restricting to systems with $0.1 \leq z \leq 0.3$. 
The final HeCS sample consists of 53 clusters from the Bright Cluster Survey \citep{ebeling} and
ROSAT-ESO FLux-LimitEd X-ray cluster survey \citep{bohringer} catalogs within the SDSS DR6 photometric footprint and with $f_{X} \geq 5*10^{-12} erg s^{-1}cm^{-2}$ and four clusters with $f_{X} \geq 3*10^{-12} erg s^{-1}cm^{-2}$.

In the case of the A2029/2033 cluster, we use the spectroscopy data from the SDSS DR7 catalog \citep{York2000,abazajian2009}
restricting to galaxies with apparent $r$ magnitude between 14.5 and 17.77. In order to identify the galaxy clusters we perform a Friends-of-friends (FoF) algorithm as described in \cite{merchan2002, merchan2005}. We use a transversal linking length corresponding to an overdensity of
$\delta \rho / \rho = 80$, a line-of-sight linking length of $V_{0}=200 km/s$, and a fiducial distance of $D_{f} =10 h^{-1}Mpc$.
We also compute a scaling factor $R$ using a galaxy luminosity function
fitted with a Schechter function with parameters ($\alpha = -1.05 \pm 0.01$, $M_{*}-5logh = -20.44 \pm 0.01$) given by \cite{blanton}.
It is worth noting that this algorithm joined A2029 with A2033, indicating that they may be physically connected.

\section{Weak-lensing analysis}
\label{sec:lensing}
 In order to obtain the projected mass density distribution and the total mass of each identified structure, we perform a similar analysis as the one presented in \citet{Gonzalez2015}. In the following Subsections, we  give the details of the criteria used for the source detection and classification,  and we describe how we obtain the shape parameters of galaxies classified as background galaxies. Finally, we obtain the two-dimensional (2D) projected density distribution and estimate the total mass of each subcluster, considering the measured ellipticities of these galaxies.

\subsection{Source identification and classification}
\label{subsec:sources}
\begin{figure*}
\centering
\includegraphics[scale=0.35]{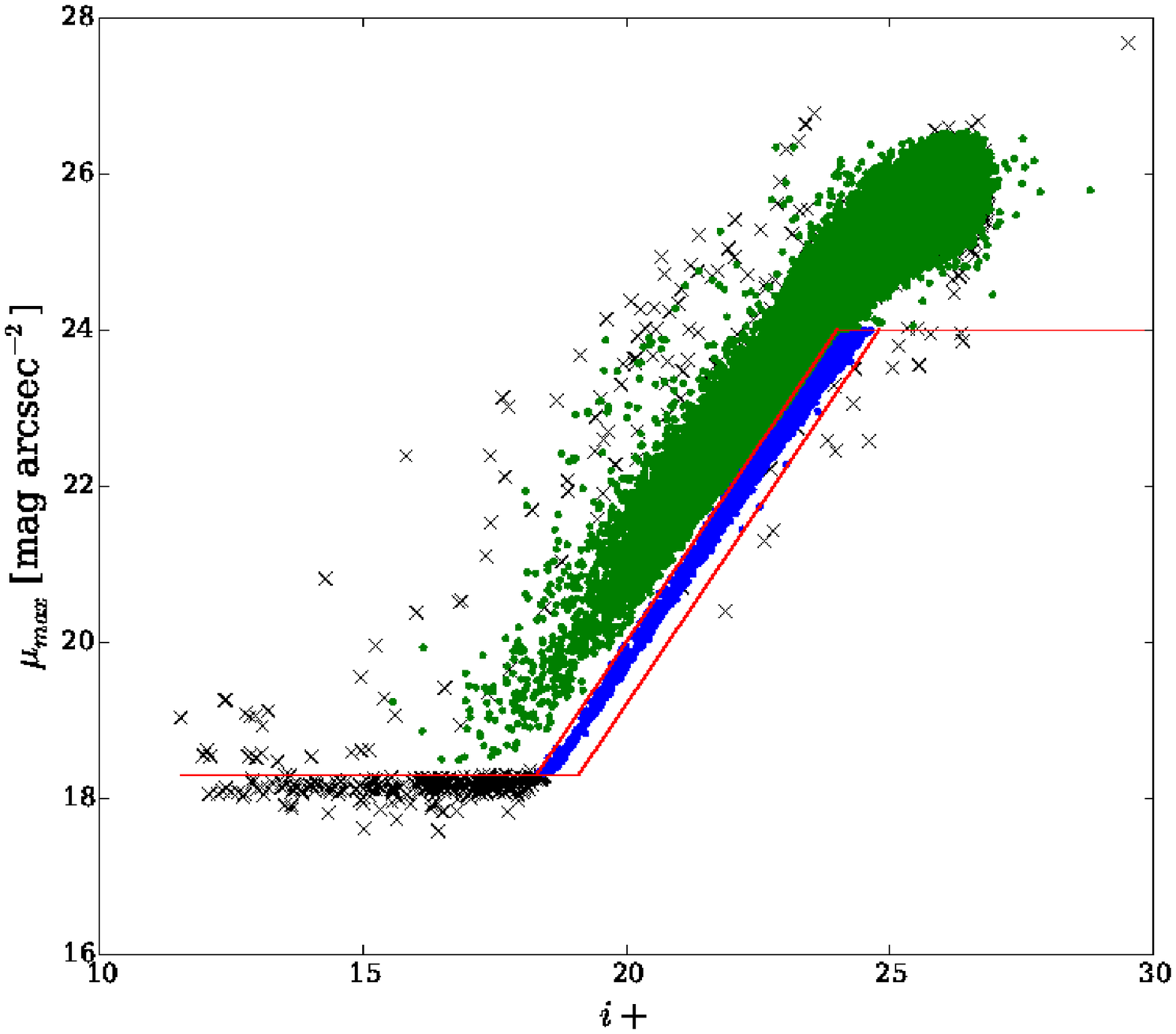}
\includegraphics[scale=0.35]{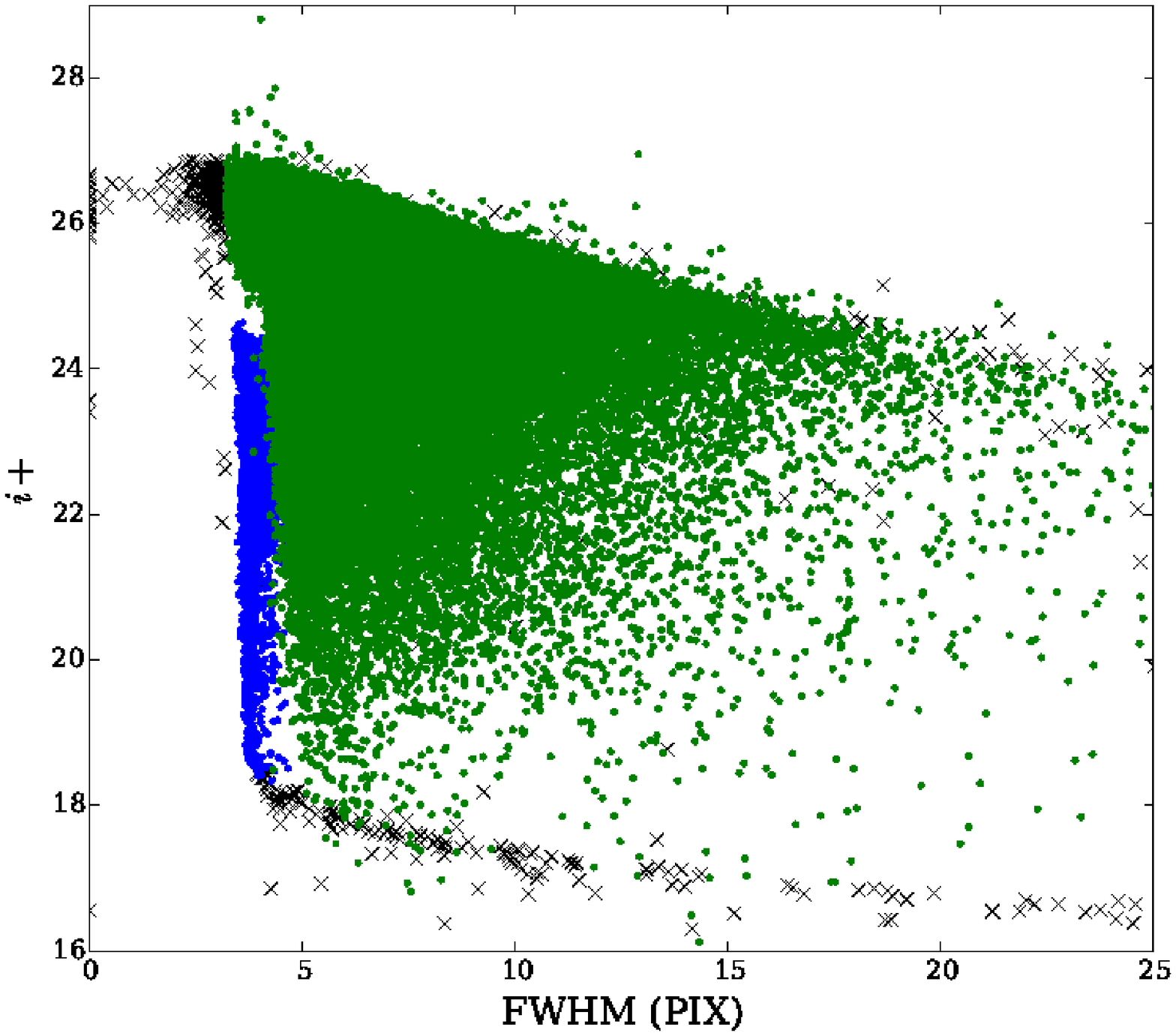}\\
\includegraphics[scale=0.35]{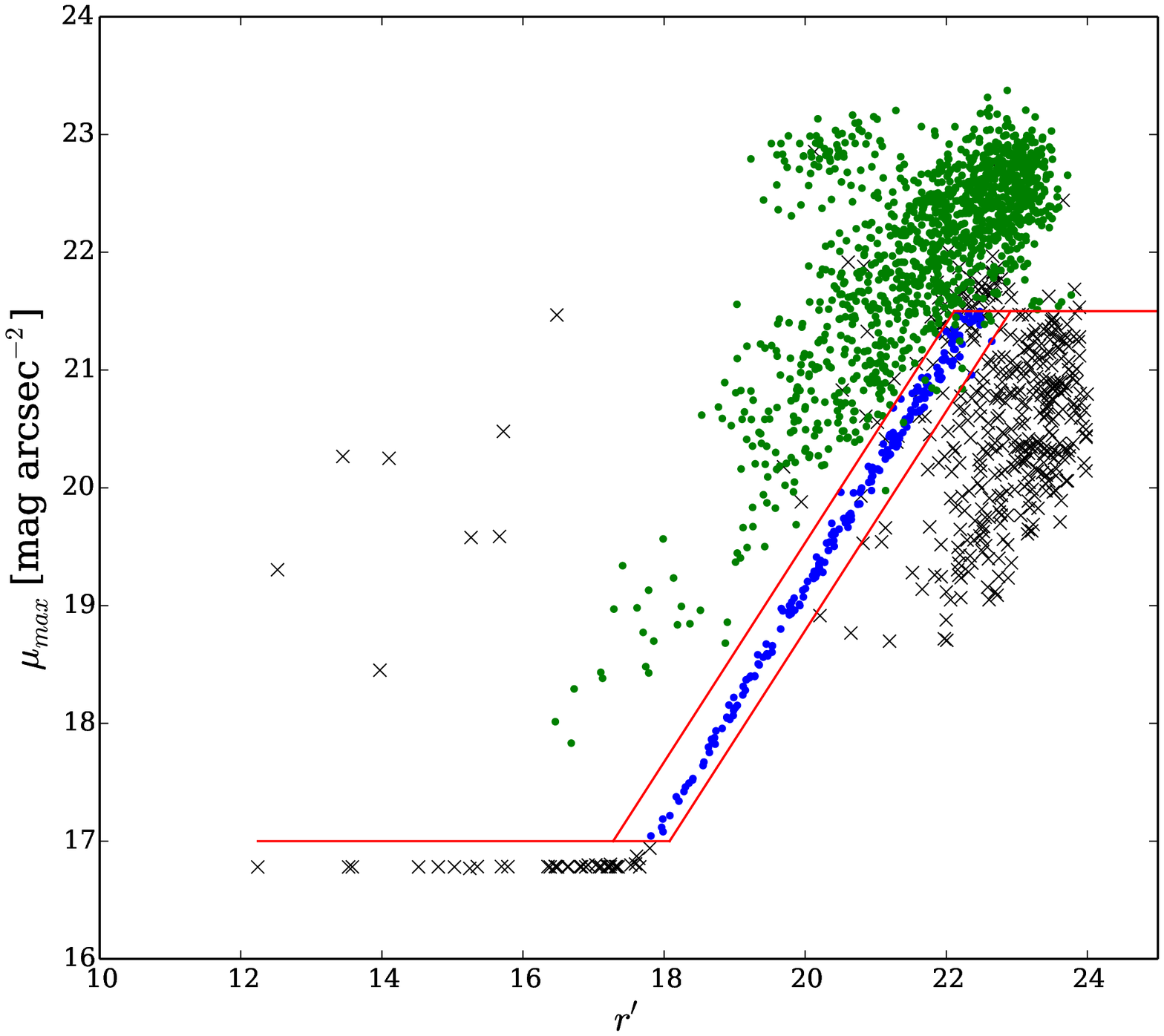}
\includegraphics[scale=0.35]{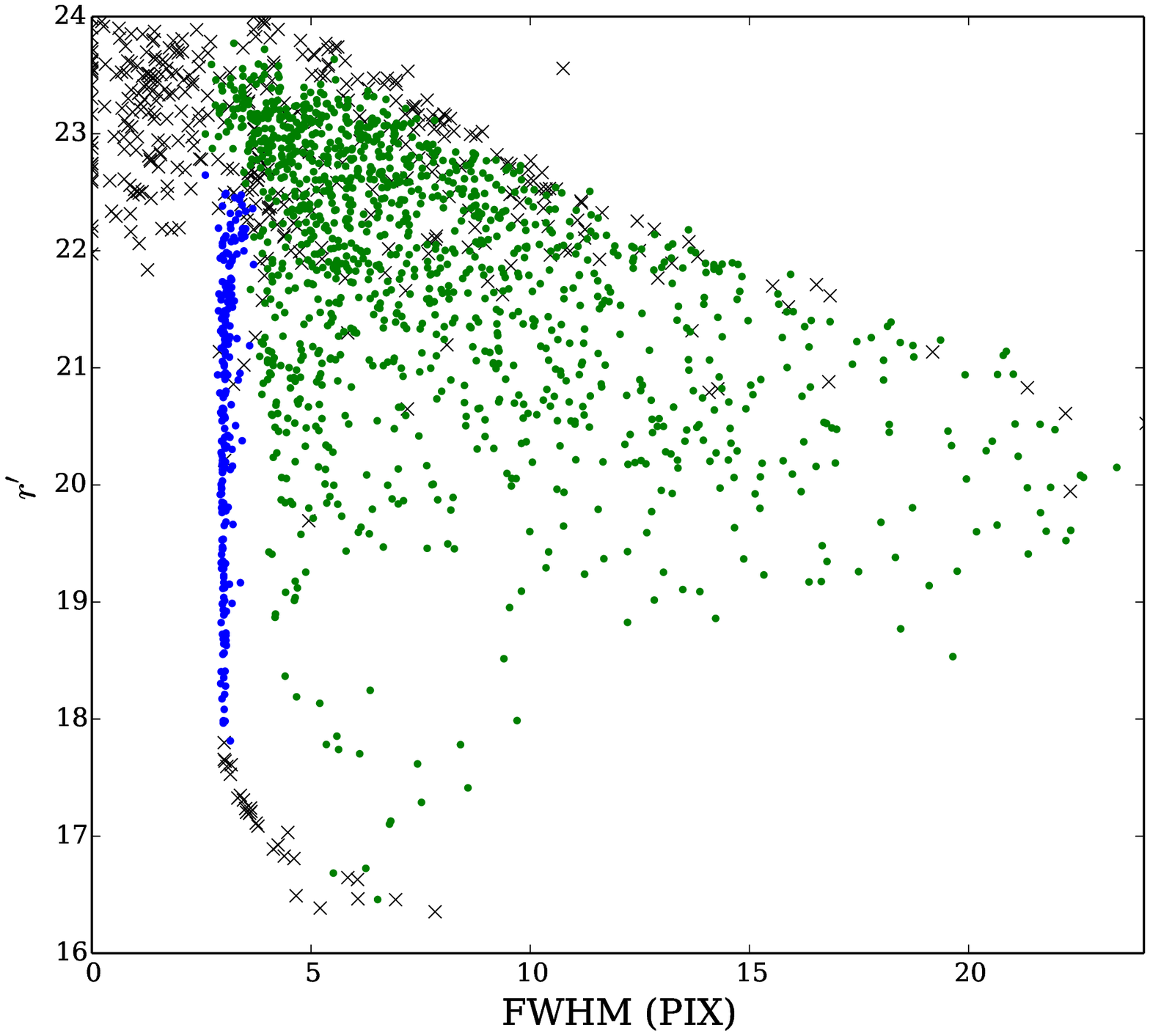}\\
      \caption{Classification of the detected sources in the SUBARU frame (\textit{upper panel}) and in one of the frames of the CFHT mosaic (\textit{lower panel}). Here stars are represented with blue points, galaxies by green points and false detections by black crosses. Left panels show the central surface brightness of the objects ($\mu_{MAX}$) as a function of the magnitude. Stars are confined in the region marked by the red lines. Right panels show the source distribution in the diagram FWHM as a function of the magnitude. As it can be seen, the FWHM of the sources classified as stars remains roughly invariant for the total range in magnitude.}
         \label{fig:source_selection}
   \end{figure*}

Source detection and photometry is carried out using \texttt{SExtractor} \citep{Bertin1996} in a two-pass mode as described in \citet{Gonzalez2015}. We detect and compute the photometry for the brightest objects by using a threshold of five times the standard deviation of  the background. This first run is made in order to compute the seeing and the saturation level of each frame. The seeing is obtained by averaging the full width at half maximum (FWHM)  of the point-like objects, according to their position in the FWHM/magnitude diagram.  The saturation level is estimated as $0.8$ times the maximum value of the FLUX\_MAX parameter, given by \texttt{SExtractor} and defined as  the peak flux above background. Then, we run \texttt{SExtractor} considering the obtained seeing and saturation level in the configuration file and with a detection threshold of $1.5$. \\
We classify the detected sources in galaxies and stars, and discard the false detections taking into account their position in the magnitude/central flux diagram, the FWHM respect to the seeing, and the stellarity index given by \texttt{SExtractor}. In Figure \ref{fig:source_selection} we show the diagrams used for the source classification. Visual inspection of the discarded sources revealed that these objects are mainly hot pixels, sources at the edges of the CCD, spikes of saturated stars, cosmic rays and overly dim sources. 

In order to perform the lensing analysis we need to identify the background galaxies which would be indeed distorted by the lensing effect. Since we do not have redshift estimations for all the identified sources, we use a photometric criterion to classify  background galaxies. Hence, we consider a galaxy as a background galaxy if its apparent magnitude satisfies $m_p < mag < m_{MAX} + 0.5$, where $mag$ is the measured magnitude in $r'$ and $i+$ bands for CFHT and SUBARU frames, respectively, $m_p$ is defined as the magnitude for which the probability that the galaxy is behind the cluster is larger than $0.7,$ and $m_{MAX}$ corresponds to the peak of the $mag$ distribution of all the identified galaxies. The upper cut ensures that we are not taking into account overly faint galaxies with higher uncertainties in the shape measurements. We obtain $m_{MAX} = 25.7$ and $m_{MAX} = 22.7$ for the SUBARU $i+$ and the CFHT $r'$ images, respectively. A discrimination by color is taken into account to discard blue galaxies which are mainly foreground galaxies that dilute the lensing signal. Therefore, we consider for the analysis only galaxies with $r'-i' > -0.5$ and $V-i+ > -1.0$ identified in the CFHT and SUBARU frames, respectively.\\
The lensing efficiency depends on the geometrical factor defined as $\beta := D_{LS}/D_S$, where $D_{LS}$ and $D_S$ are the angular diameter distances from the lens to the source and from the observer to the source, respectively. To estimate $m_p$ and $\langle \beta \rangle$ (where $\langle ...\rangle$ express the average over the considered background galaxies for the lensing estimator) we use catalogs of photometric redshifts. In the case of the lensing analysis performed with CFHT frames we use the \citet{Coupon2009} photometric catalog, based on the public release Deep Field 1 of the Canada–France–Hawaii Telescope Legacy Survey (CFHTLS Deep1), which has an $80 \%$ completeness limit of $m_{r'} \approx 26.$  and  covers a sky region of roughly 1\,deg$^{2}$ . On the other hand, for the analysis with SUBARU frames we use the catalog of photometric redshifts given by \citet{Laigle2016} which contains precise photometric redshifts over the 2deg$^2$ COSMOS field and with a limiting magnitude $m_{i+}=26.2$. We compute the fraction of galaxies with $z >  z_{cluster}$ in magnitude bins of 0.25 $mag$ and then we chose $m_{P}$ as the lowest magnitude for which the fraction of galaxies was greater than $0.7$, obtaining $m_{P}=18.2$ and $m_{P}=18.5$ for $i+$ and $r'$ images, respectively. Then we apply the photometric selection criteria ($m_{P} < m_{mag} < m_{MAX} + 0.5$) to the catalog and we compute $\beta$ for the whole distribution of galaxies. To take into account the contamination by foreground galaxies given our selection criteria, we set $\beta(z_{phot} < z_{cluster}) = 0$  which outbalances the dilution of the shear signal by these unlensed galaxies.\\
To estimate the error in $\langle \beta \rangle$ regarding the cosmic variance, we divide CFHTLS Deep1 and COSMOS fields into 25 and 64 non-overlapping areas of $\sim$\,144\,arcmin$^2$ and $\sim$\,160\,arcmin$^2$, respectively. Then we compute $\langle \beta \rangle$ for each area considering the redshift of the analyzed clusters according to the used images. The uncertainties in $\langle \beta \rangle$ due to cosmic variance are estimated according to the scatter among the values for each area, obtaining  $\sim 0.012$ for CFHTLS DEEP1 field and $\sim 0.010$ for COSMOS field. These uncertainties were taken into account in the error estimation of the fitted parameters, and propagated to the resulting system masses.\\
In order to take into account the contamination of foreground galaxies in the catalog, we compute for each galaxy the probability that it is behind the cluster. This probability is computed using the described photometric catalogs considering the fraction of galaxies with $z >  z_{cluster}$ for each bin in magnitude, $mag$, and color (\textit{V - i+} and \textit{r' - i'}). Hence, given the magnitude and the color of each galaxy, we assign to it a weight, \textit{w}, as the fraction of galaxies with $z >  z_{cluster}$ in that bin. This weight is applied to compute the mass and the 2D density distribution.

\subsection{Shape measurements}
\label{subsec:shape}

\begin{figure*}
\centering
\includegraphics[scale=0.4]{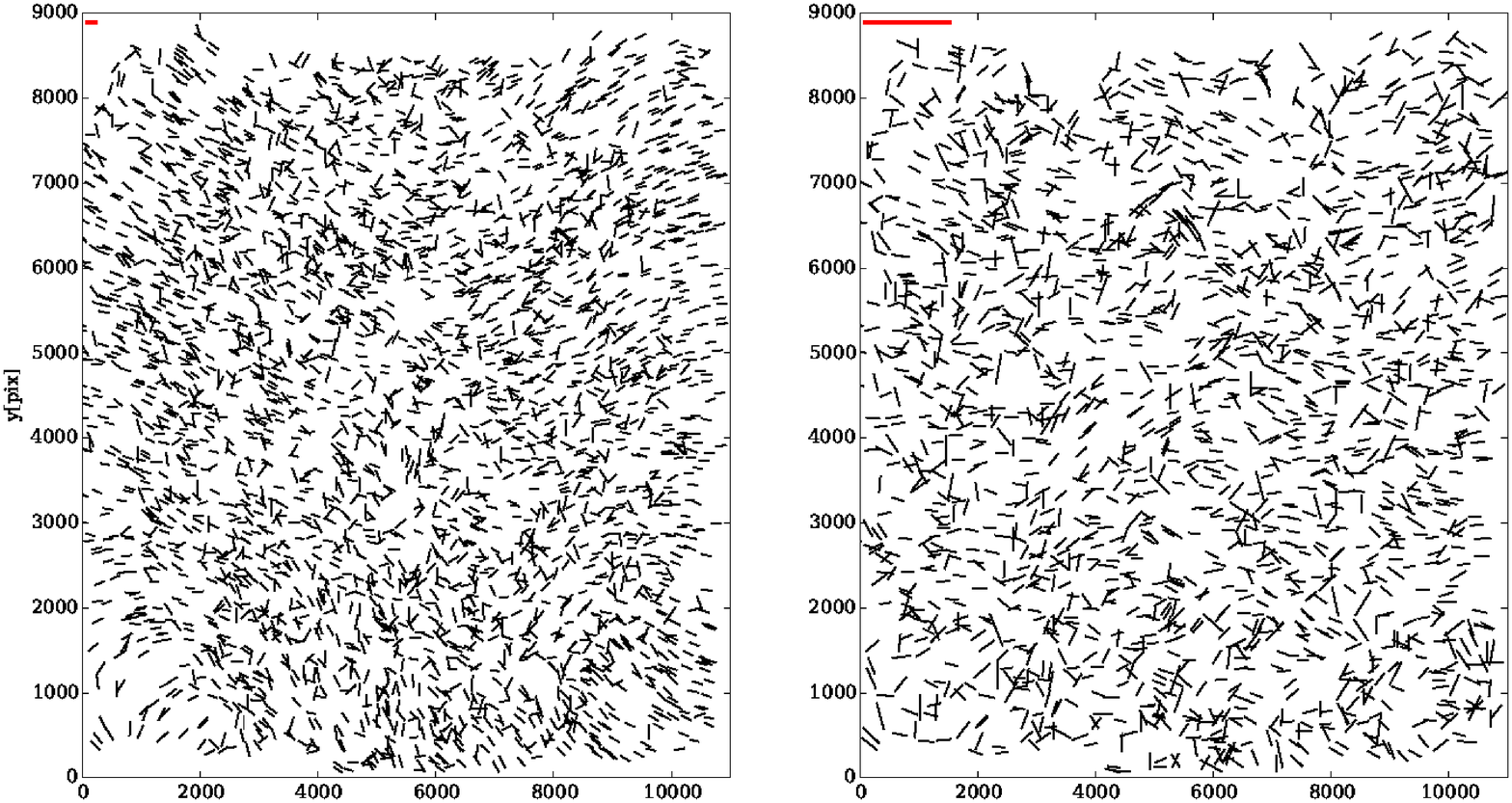}\\
\includegraphics[scale=0.4]{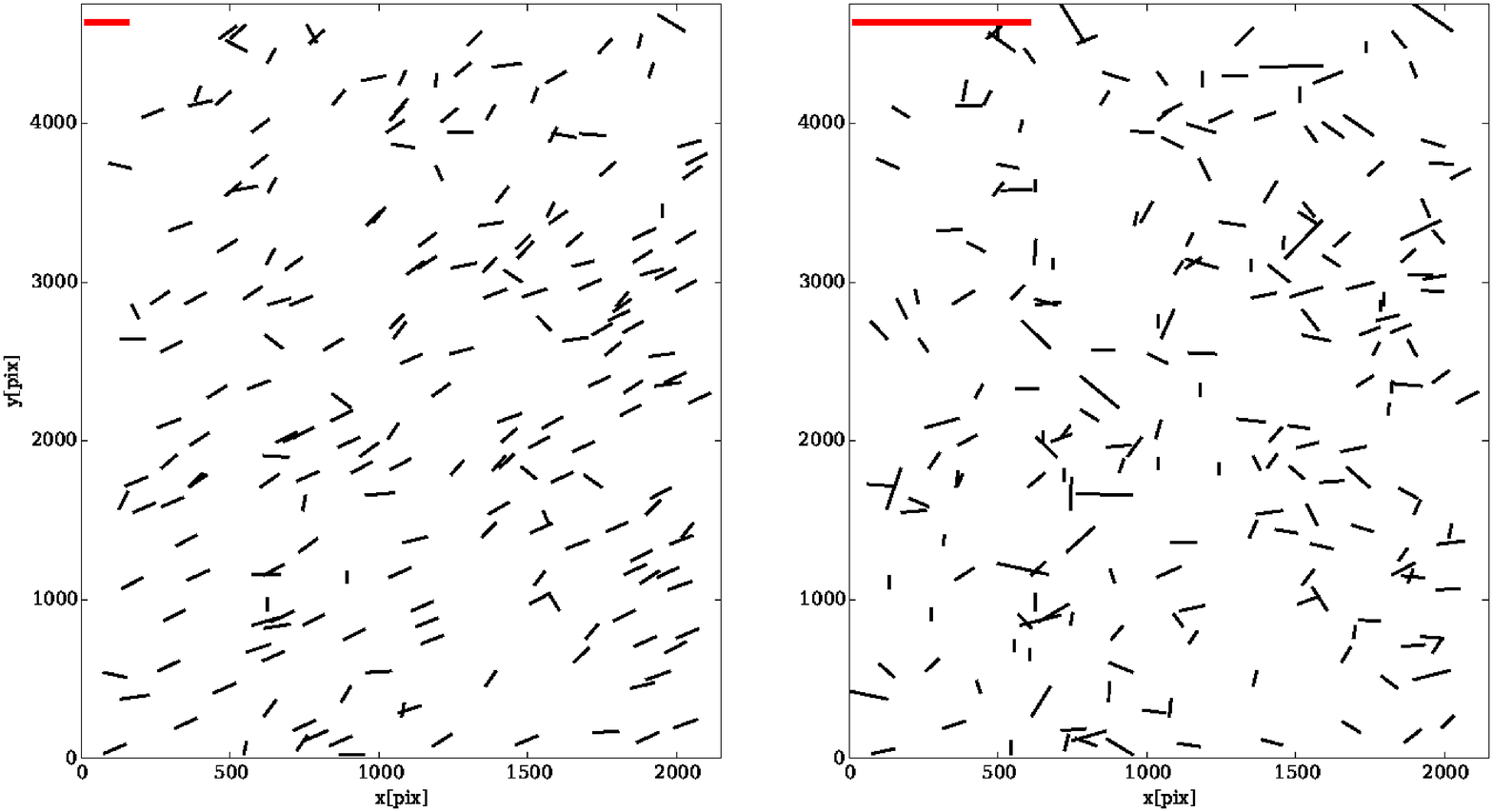}

      \caption{PSF treatment applied to stars in the SUBARU frame (\textit{upper panel}) and in one of the frames of the CFHT mosaic (\textit{lower panel}). Semi-major axis distribution in the CCD, of the sources classified as stars, before (left) and after (right) the PSF
deconvolution. For a better visualization of the plot, semi-major axes have a different scale than the CCD axes. Semi-major axis scale is given by the red thicker segment in the upper left-hand corner and corresponds to 3 pixels. We highlight that the semi-major axis is more randomly distributed and is significantly smaller after taking into account the PSF.}
         \label{fig:PSF_map}
   \end{figure*}

Gravitational lensing effects are characterized by an isotropic stretching called convergence, $\kappa$, and an anisotropic distortion given by the complex-value lensing shear, $\gamma = \gamma_1 + i \gamma_2$.  Using the second  derivative  of  the  projected  gravitational  potential  to express the shear and convergence, one can show that for a lens with a circular-symmetric projected mass distribution, the tangential component of the shear, $\gamma_T$, is related to the convergence through \cite{Bartelmann1995}:
\begin{equation}
\gamma_T (r) = \bar{\kappa}(<r)-\langle \kappa \rangle (r)
,\end{equation}
where $\bar{\kappa}(<r)$ and $\langle \kappa \rangle (r)$ are the convergence averaged over the disc and circle of radius $r$, respectively. On the other hand, the cross component of the shear, $\gamma_\times$ defined as the component tilted at $\pi/4$ relative to the tangential component, should be exactly zero.\\
If lensing is weak ($\kappa \ll 1$), the image of a circular source appears elliptical with $a$ and $b$ as major and minor semi axes, respectively. The induced ellipticity could be directly related to the shear, $\gamma \approx e$. Here we define the ellipticity as a complex number, $e = e_1 + i e_2$, with magnitude $\mid e \mid = (a-b)/(a+b)$ and orientation angle determined by the direction of the major elliptical axis. If the source has an intrinsic ellipticity $e_s$ the observed ellipticity would be $e \sim e_s + \gamma$ \citep{Bartelmann2001}. If we consider many sources with intrinsic ellipticities randomly orientated so that $\langle e_s \rangle = 0$, the ensemble average ellipticity after lensing gives an unbiased estimate of the shear: $\langle e \rangle \approx \gamma$. Therefore, we can estimate the shear by averaging the shape of background sources that would be affected by the lensing effect.\\
In order to perform the lensing analysis and, in turn, to estimate the projected masses of the galaxy systems, it is necessary to obtain the shape parameters of background galaxies. It is important to take into account the roundness effects due to the atmosphere presence, as well as the distortion caused by the telescope effects. All of these are considered by the \textit{Point Spread Function} (PSF), which is convolved with the true galaxy intensity light distribution.\\
To obtain the shape parameters we use \texttt{im2shape} \citep{Bridle2002}, which models the galaxies as a sum of Gaussians convolved with a PSF, which is, in turn, a sum of Gaussians. For simplicity both the PSF and the object are modeled with a single elliptical Gaussian profile. To estimate the PSF at the position of each galaxy we average the shape parameters of the five closest sources classified as stars, since they are considered point-like sources \citep[this procedure is described in ][]{Gonzalez2015}. To check our PSF correction we obtain the shape parameters of the stars with and without considering the PSF. In Figure \ref{fig:PSF_map} we show the semi-major axis distribution of the stars before and after considering the computed PSF for obtaining the shape parameters. As can be seen from the Figure, when we consider the estimated PSF at the position of each star, the semi-major axes are more randomly distributed and their lengths are substantially smaller, according to the distribution that would be presented by point-like sources. 

To compute the lensing masses and the 2D density distributions we only kept the galaxies classified as background galaxies and with FWHM > 6 pix, to ensure galaxies with a good pixel sampling and, therefore, with better-shaped parameter estimations. Also, we discard galaxies with $\sigma_e > 0.2$, where $\sigma_e$ is defined as the error in the measured ellipticity and is computed as $\sigma_e = \sqrt{\sigma_{e_1}^2 + \sigma_{e_2}^2}$, where $\sigma_{e_1}$ and $\sigma_{e_2}$ are the errors in the ellipticity components, $e_1$ and $e_2$, respectively, provided by \texttt{im2shape}. With these criteria we obtain a density of background galaxies of $\sim$ 17.4 galaxies/arcmin$^2$ for SUBARU image and $\sim 3.5$ galaxies/arcmin$^2$ for CFHT mosaic image.

\subsection{Individual masses}

Once we obtain the shape parameters of the background galaxies we can compute the shear profile which could be fitted by assuming a density distribution for each structure identified in the analyzed systems. Spherical symmetry implies that the average in annular bins of the tangential component ellipticity, $e_T$, of the lensed galaxies traces the reduced shear. On the other hand, the average in annular bins of the component tilted at $\pi/4$ relative to the tangential component, $e_\times$, should be exactly zero for the case of perfect symmetry \citep[e.g., ][, Sect. 4]{Bartelmann2001}. Since galaxies have an intrinsic ellipticity, the error in the shear estimator $\langle \gamma(r) \rangle$, obtained by averaging the tangential ellipticity component of the $N$ galaxies at a distance $r\pm \delta r$ from the center considered, $\langle e_T(r) \rangle$, would be \citep{Schneider2000}:
\begin{equation} \label{eq:error}
\sigma_\gamma \approx \frac{\sigma_\epsilon}{\sqrt{N}}
,\end{equation}
where $\sigma_{\epsilon}$ is the dispersion of the intrinsic ellipticity distribution \citep[$\sigma_{\epsilon} \approx 0.3$,][which is fixed at this value for the rest of the analysis]{Leauthaud2007}. The brightest galaxy member identified from each structure, according to the membership presented in the catalog of \cite{delosRios2016}, is considered as the center to build the profiles. Shear profiles are computed using non-overlapping logarithmic annuli, in order to have a similar signal-to-noise ratio (S/N) in each annuli. To obtain the shear estimator, $\langle \gamma(r) \rangle$, we compute a weighted average of tangential ellipticity components of the $N$ galaxies in the bin center at a radius $r$. The weight considered is the probability computed for each galaxy to be behind the cluster, as was described in Subsection \ref{subsec:sources}. We test different annuli sizes but the final mass results do not show a strong dependence on this parameter. We fix the size for the one for which we obtain the lowest $\chi^2$ value for the fitted parameters. Profiles are fitted from the inner part where the signal becomes significantly positive, to reduce the impact of miss-centering, up to $\sim 1.5 Mpc$ to avoid the region where the signal relating the companion structure becomes significant. \\
Two mass models are used to fit the resultant shear profiles: a singular isothermal sphere (SIS) and a NFW profile \citep{Navarro1997}. The SIS profile is the simplest density model for describing a relaxed massive sphere with a constant value for the isotropic one dimensional velocity dispersion, $\sigma_V$. This is mainly described by the density distribution:
\begin{equation}\label{eq:SIS_density}
\rho(r) =  \dfrac{\sigma_{V}^{2}}{2 \pi G r^{2}}
\end{equation}
where $G$ is the gravitational constant. This model corresponds to a distribution of self-gravitating particles where the velocity distribution at all radii is a Maxwellian with one dimensional velocity dispersion, $\sigma_{V}$. From this equation, we can get the critical Einstein radius for the source sample as:
\begin{equation}\label{eq:SIS}
\theta_{E} = \dfrac{4 \pi \sigma_{V}^{2}}{c^{2}} {\langle \beta \rangle}
,\end{equation}
in terms of which one obtains:
\begin{equation}
\kappa ({\theta}) = \gamma({\theta}) = \dfrac{\theta_{E}}{2 \theta}
,\end{equation}
where $\theta$ is the angular distance to the cluster center. Hence, fitting the shear for a different radii, we can estimate the Einstein radius, and from that, we can obtain an estimation of the mass M$_{200}$ as \citep{Leonard2010}:
\begin{equation}\label{eq:MSIS}
M_{200} =  \dfrac{2 \sigma_{V}^{3} }{\sqrt{50} G H(z)} 
,\end{equation} 
where $H(z)$ is the redshift dependent Hubble parameter.

The NFW profile is derived from fitting the density profile
of numerical simulations of cold dark-matter halos \citep{Navarro1997}. This profile depends on two parameters, the radius, $R_{200}$, that encloses a mean density equal to 200 times the critical density ($\rho_{crit} \equiv 3 H^{2}(z)/8 \pi G$), and a dimensionless concentration parameter, $c_{200}$:
\begin{equation}
\label{eq:NFW_density}
\rho(r) =  \dfrac{\rho_c \delta_c}{(r/r_{s})(1+r/r_{s})^{2}} 
\end{equation}
where $r_{s}$ is the scale radius, $r_{s} = R_{200}/c_{200}$ and $\delta_{c}$ is the cha\-rac\-te\-ris\-tic overdensity of the halo,
\begin{equation}
\label{eq:NFW_delta}
\delta_{c} = \frac{200}{3} \dfrac{c_{200}^{3}}{\ln(1+c_{200})-c_{200}/(1+c_{200})}  
\end{equation}
We use the lensing equation for the spherical NFW density profile from \citet{Wright2000}. If we fit the shear for different radius we can obtain an estimation of the pa\-ra\-me\-ters $c_{200}$ and $R_{200}$. Once we obtain $R_{200}$ we can compute the $M_{200}$ mass. Nevertheless, there is a well-known degeneracy between the parameters $R_{200}$ and $c_{200}$ when fitting the shear profile in the weak lensing regime. This is due to the lack of information on the mass distribution near the cluster center and only a combination of strong and weak lensing can raise it and provide useful constraints on the concentration parameter. Due to the lack of strong lensing modeling in our sample, we follow \citet{Uitert2012,Kettula2015} and \citet{Pereira2017}, by fixing the concentration parameter according to the relation $c_{200}(M_{200},z)$ given by \citet{Duffy2011}: 
\begin{equation}
c_{200}=5.71(M/2 \times 10^{12} h^{-1})^{-0.084}(1+z)^{-0.47}
.\end{equation}
The particular choice of this relation does not have a significant impact on the final mass values, with uncertainties dominated by the noise of the shear profiles. Thus we fit the profile with only one free parameter: $R_{200}$.\\
To derive the parameters of each mass model profile we perform a standard $\chi^{2}$ minimization:
\begin{large}
\begin{equation}
\chi^{2} = \sum^{N}_{i} \dfrac{(\langle \gamma(r_{i})  \rangle - \gamma(r_{i},p))^{2}}{\sigma^{2}_\gamma(r_{i})}
,\end{equation}
\end{large}\\
where the sum runs over the $N$ radial bins of the profile and the model prediction. $p$ refers to either $\sigma_{V}$ for the SIS profile, or $R_{200}$ in the case of the NFW model. Errors in the best-fitting parameters are computed according to the variance of the parameter estimate.
 
 \subsection{Two-dimensional density distribution}

We obtain the 2D projected density distribution for the two analyzed galaxy systems by using the \texttt{LensEnt2} code \citep{Bridle1998,Marshall2002}. This code applies maximum-entropy method algorithm for reconstructing the 2D density distribution from the two ellipticity components, $e_1$ and $e_2$, of the background galaxies and their respective uncertainties. We consider $\sigma_\epsilon/w$ as the uncertainty for both components, where $w$ is the assigned weight to each galaxy, as was explained in Subsection \ref{subsec:sources}. It is expected that galaxy clusters have smooth, extended projected mass distributions, therefore, an Intrinsic Correlation Function (ICF) is included in the analysis, so the physical projected density distribution, $\Sigma$, is expressed as the convolution with a broad kernel, given by the ICF. In other words, the ICF smooths the resultant distribution. We adopt a circularly symmetric Gaussian as the ICF with a width $\omega$. This width could be selected to maximize the Bayesian evidence, nevertheless this could lead to a sub-estimation of the substructure in the field.  Hence, a visual inspection of the obtained maps and a comparison with other constraints (X-ray contours, galaxy member distribution and the position of the BCG) could be important in order to select an adequate value. Taking this into account we obtain the density distributions for different widths, $\omega$, and then we fix this value considering the resultant Bayesian evidence and according to a visual inspection of the reconstructed maps.
We fix $\omega$ at 550\,kpc and 410\,kpc for A1204 and A2029/2033, respectively.\\
Finally to obtain the projected density distribution, $\Sigma$, it is necessary to compute the critical density defined as:
\begin{equation}
\Sigma_c=\frac{c^2}{4 \pi G} \frac{1}{D_L \langle \beta \rangle}
,\end{equation}
where we adopt the parameters $D_L$ and $\langle \beta \rangle$ for the primary component.\\
The code also provides an error map according to the covariance matrix of the density reconstruction. We consider as the error in our 2D density distribution the mean of the values in the error map. Contours in the 2D distribution are built from a 3$\sigma$ confidence level, while the distribution is shown from a $2\sigma$ confidence level.

\section{Dynamical analysis}
\label{sec:dynamic}

In a previous work \cite{delosRios2016} studied the dynamical properties of the clusters A2029/2033 and A1204 using photometric 
and spectroscopic information. They estimated some parameters that correlate with the dynamical status of the clusters and then, 
by using machine learning algorithms, classified them as candidates for merging systems \citep[see Table\,1 in ][]{delosRios2016}.
In this Section, using the publicly available Merging Systems Identification Algorithm \footnote{\href{https://martindelosrios.netlify.com}{https://martindelosrios.netlify.com}}, we extend their analysis and describe the dynamical state of A1204 and A2029/2033.\\
For the A1204 cluster we study its dynamical status with the merging system identification algorithm described in \citet{delosRios2016} and we find that this is a merging cluster candidate. In order to study the stability of our result, we apply the same technique but deleting some randomly selected galaxies and then we analyze how the classification changes with the percentage of dropped galaxies. In particular, we perform 100 realizations for each value of completeness ($99\%$, $95\%$ and $90\%$). In all cases, we classify the system as a merging cluster in $\sim 50\%$ of the realizations, suggesting that the classification is not stable, therefore if there is an interaction between the components, it will not be strong. On the other hand, we also apply the identification algorithm separately to the two components identified by the algorithm in the first study. We found that each individual component is classified as not being in the process of merging, indicating that this is not a case of a multiple merger.\\
In the case of the A2029/2033 cluster, we analyze the cluster identified by the FoF algorithm (i.e., Abell 2029 and Abell 2033 as a single cluster) with the Merging Systems Identification
Algorithm. We find this system to be a merging cluster candidate and that A2029 and A2033 are the two substructures that are interacting. As in the case of A1204, we study the stability of the classification by reclassifying the dynamical status of the FoF cluster in random realizations dropping a percentage of member galaxies. In all the realizations, the algorithm classified the cluster as a merging system, hence, the classification is stable. We also analyze each system (A2029 and A2033) separately applying the Merging Systems Identification
Algorithm and we obtain that each cluster is classified as not being in the process of merging. This is in agreement with previous classifications of the individual clusters as relaxed systems \citep{wen2013,mantz2016}.\\

\section{Results and discussion}
\label{sec:results}

\begin{figure}
\centering
\includegraphics[scale=0.45]{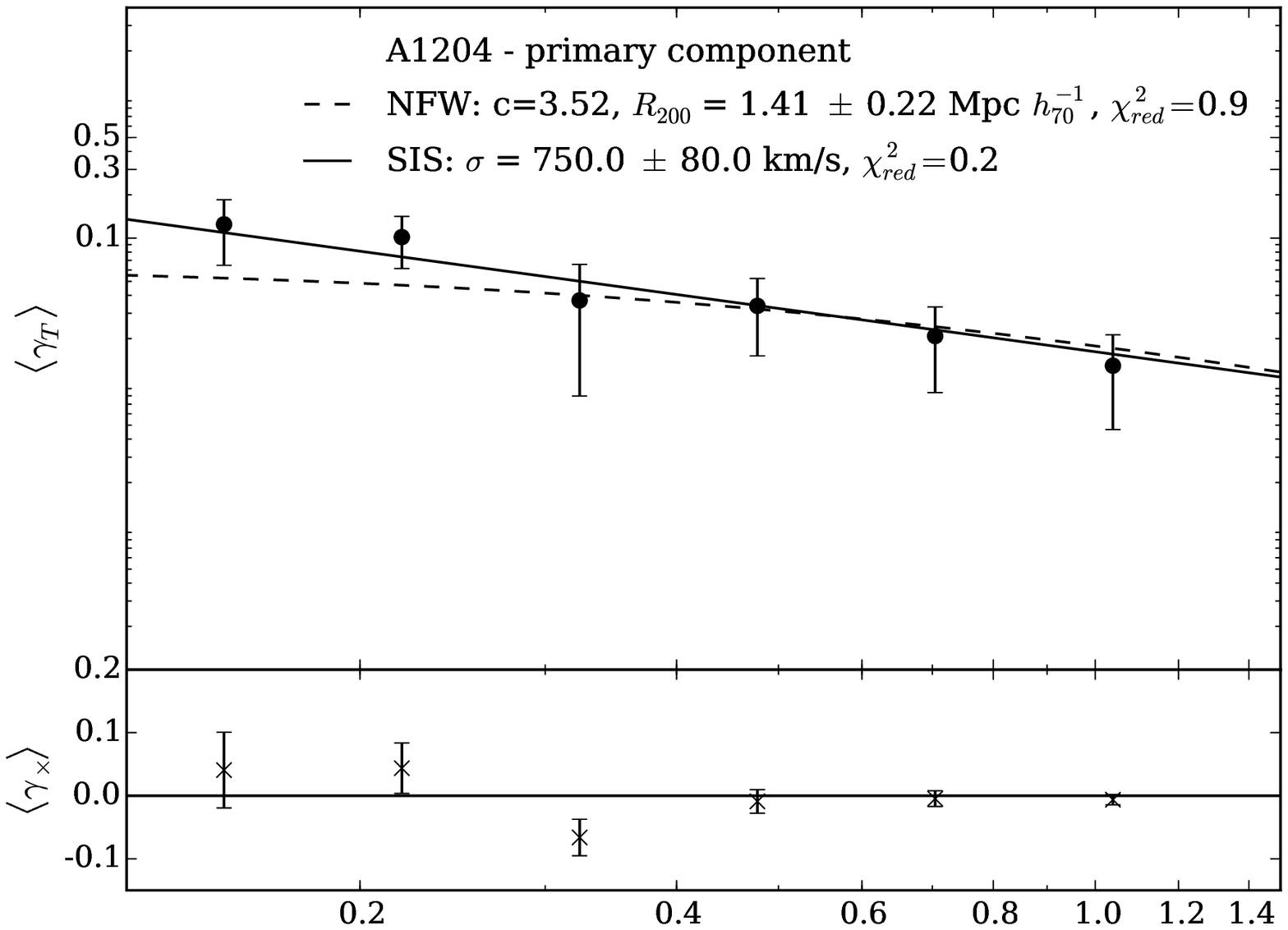}
\includegraphics[scale=0.45]{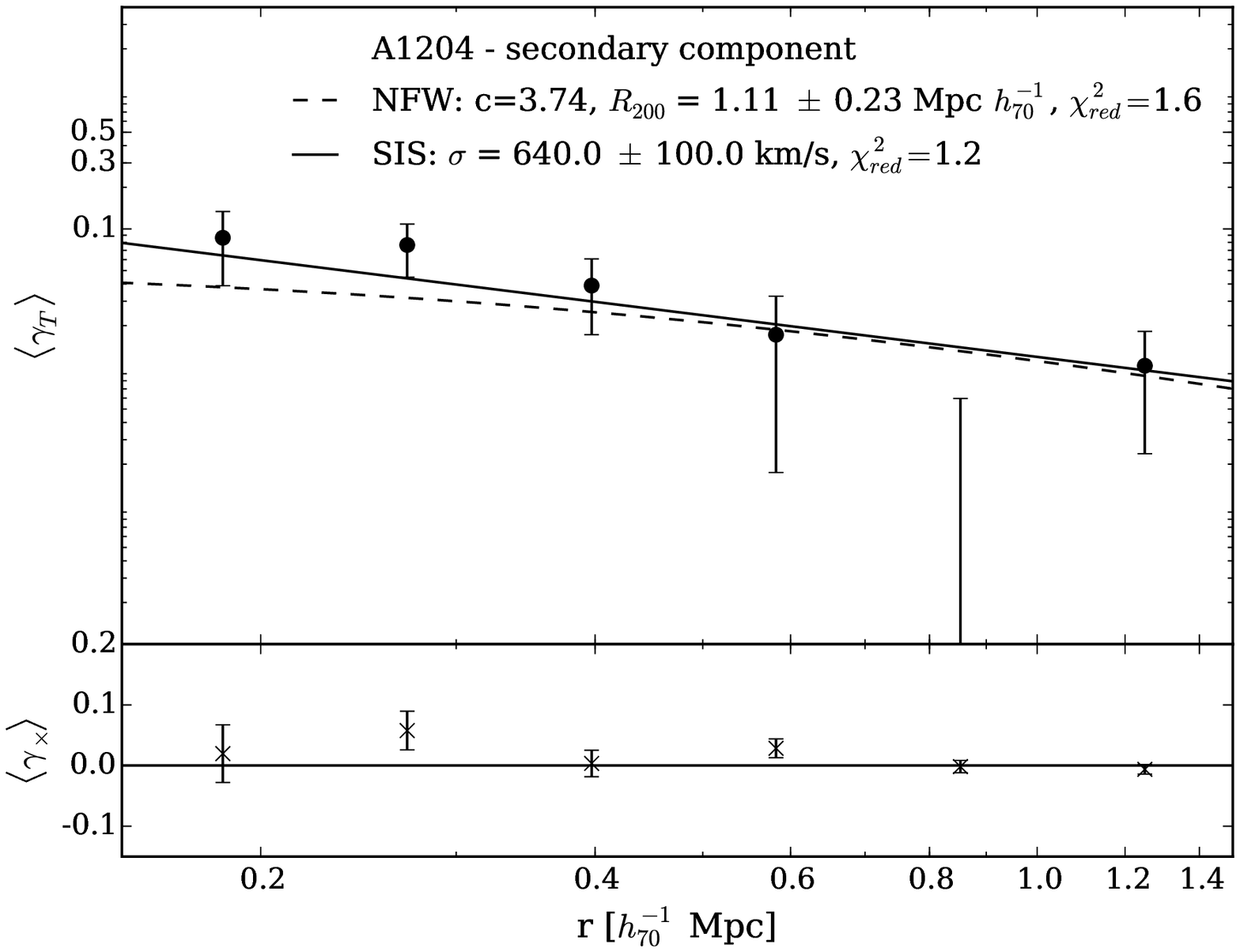}\\
      \caption{Shear radial profiles as a function of cluster-centric projected distance for the primary (\textit{upper panel}) and the secondary component (\textit{bottom panel}) of A1204, identified by \citet{delosRios2016}. The solid and the dashed lines represent the best fit of SIS and NFW profiles, respectively, with the fitted parameters given in the right-upper corner. The points and crossings show the tangential and cross ellipticity components of the selected background galaxies, averaged in annular bins, respectively. Error bars are computed according to Equation \ref{eq:error}.}
         \label{fig:shear_profile_A1204}
\end{figure}

\begin{figure}
\centering
\includegraphics[scale=0.45]{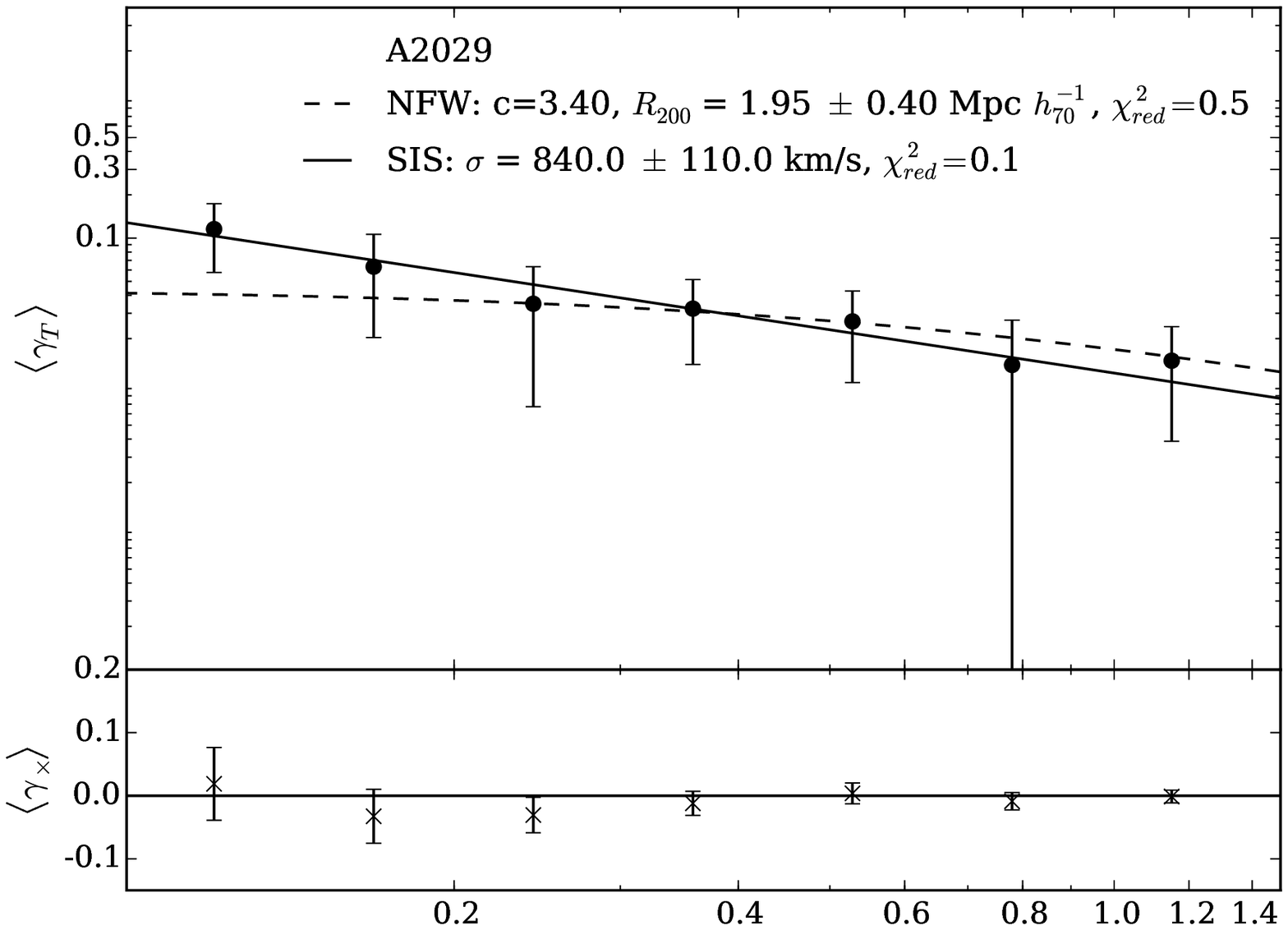}\\
\includegraphics[scale=0.45]{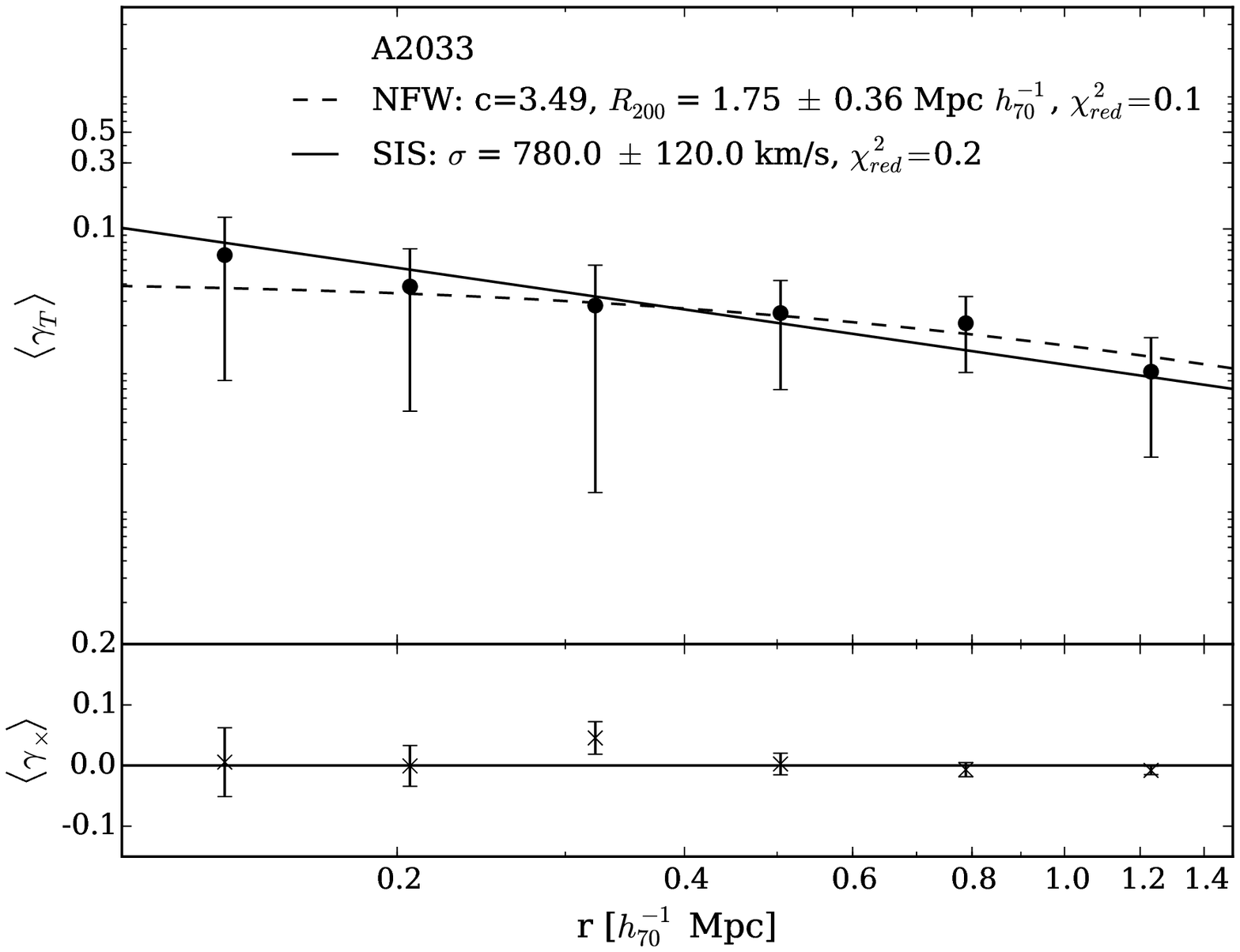}\\
\caption{As in Figure \ref{fig:shear_profile_A1204} but for A2029 (\textit{upper panel}) and A2033 (\textit{bottom panel}).
      }
         \label{fig:shear_profile_A2029/2033}
\end{figure}

\begin{figure*}
\centering
\includegraphics[scale=0.4]{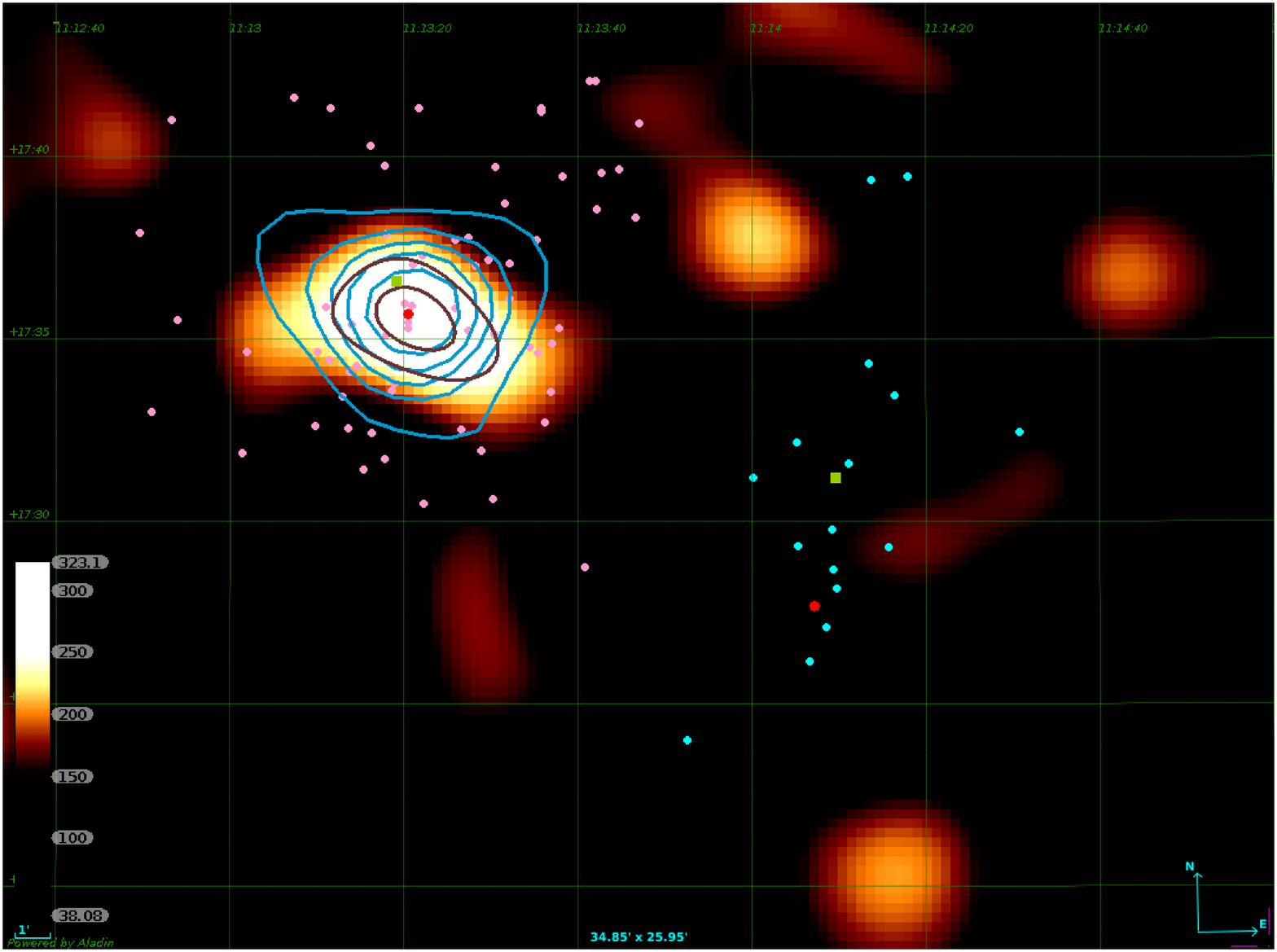}
\includegraphics[scale=0.4]{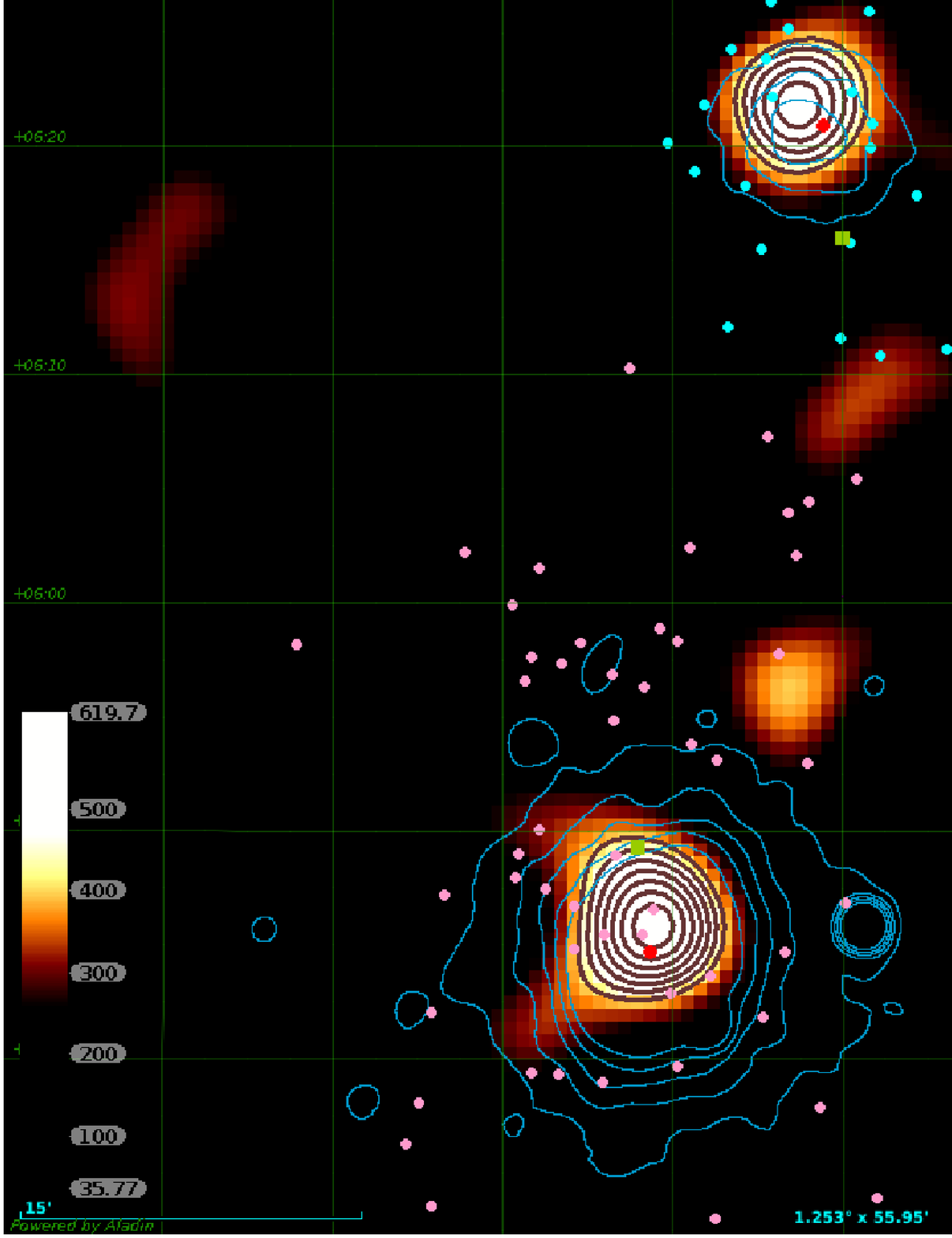}\\

      \caption{Projected density distribution in the field A1204 (\textit{upper panel}) and in the field of A2029 and A2033 (\textit{lower panel}) obtained with the weak lensing analysis. The scale, marked at bottom left, is given in h$_{70} M_\odot/pc^2$. Red contours corresponds to a projected density above $3\sigma$ significance level (250 h$_{70} M_\odot/pc^2$ for A1204 and 400 h$_{70} M_\odot/pc^2$ for A2029/A2033) with steps of $50 M_\odot/pc^2$. X-ray contours are plotted in blue, the contour
levels are (3, 5, 7, 9 and 12) times the rms noise. Red dots and green squares are the BCGs positions and the dynamical centres, respectively. Pink and light-blue points are the positions of the galaxies classified as members, for the primary and secondary component of A1204 and for A2029 and A2033, respectively.}
         \label{fig:mass-map}
\end{figure*}

In this Section we present the results of the lensing analysis, together with the X-ray and dynamical characterization of the systems. In Table \ref{tab:results} we show the obtained lensing masses for each structure according to the fitted shear profiles (Figs. \ref{fig:shear_profile_A1204} and \ref{fig:shear_profile_A2029/2033}). In order to characterize
the gas distribution, we compute brightness contour maps at
(3, 5, 7, 9, 12) times the background and smoothed over 4 pixels. The derived 2D projected density distributions together with X-ray contours are shown in Figure \ref{fig:mass-map}.

\begin{table*}

\caption{Main results of the weak lensing analysis.}\label{tab:results}

\begin{tabular}{@{}crrrrccrccr@{}}
\hline
\rule{0pt}{1.05em}%
   Galaxy system  &  $\alpha$ & $\delta$  & $z$ & $\langle\beta\rangle$& \multicolumn{2}{c}{SIS} & \multicolumn{3}{c}{NFW}   \\
   Id. & (J2000) & (J2000) &  &   & $\sigma_{V}$ & $M_{200}$ & $c_{200}$ &$R_{200}$  & $M_{200}$   \\
       &         &         &  &   & km\,s$^{-1}$ & h$_{70}^{-1} 10^{14} M_{\odot}$ & & h$_{70}^{-1}$ Mpc &  h$_{70}^{-1} 10^{14} M_{\odot}$ \\
 \hline
\rule{0pt}{1.05em}%
A1204 I  &  11h 13m     20.5s   &   +17$^\circ$  35'  41.0''    &   0.1703 &0.689 & 750   $\pm$ 80  &  3.7 $\pm$  1.2  & 3.52  & 1.41 $\pm$ 0.22 &  3.8 $\pm$  1.8 \\ 
A1204 II  &  11h 14m 07.2s      &   +17$^\circ$  27'  41.0''    &  0.1705 &0.689 &   640   $\pm$ 100  &  2.3 $\pm$  1.1  & 3.52 & 1.11 $\pm$ 0.23 &  1.9 $\pm$  1.2 \\ 
A2029  &  15h 10m       56.1s   &       +05$^\circ$      44' 41.7''     & 0.0775 &  0.778  & 840   $\pm$ 110  &    5.3 $\pm$    2.0  & 3.40 & 1.95 $\pm$ 0.40 &  9.1 $\pm$  5.6  \\ 
A2033  &  15h 11m       26.5s   &   +06$^\circ$  20'  56.9''    & 0.0822 &  0.766 & 780   $\pm$ 120  &  4.3 $\pm$  2.1  & 3.49 & 1.75 $\pm$ 0.36 &  6.5 $\pm$  4.0 \\ 
\hline         
\end{tabular}
\medskip
\begin{flushleft}
\textbf{Notes.} Columns: (1) shows the cluster identification; columns (2), (3) and (4): the coordinates of the center and the redshift adopted for the lensing analysis; column (4) and (5): the $m_P$ and $m_{max}$ magnitudes considered for the galaxy background selection (see Subsection \ref{subsec:sources}); column (6): the geometrical factor; columns (7) and (8): the results from the SIS profile fit, the velocity dispersion and M$_{200}$ (see Equations\,\ref{eq:SIS} and \ref{eq:MSIS}); columns (9), (10) and (11): the results from the NFW profile fit, the adopted $c_{200}, $R$_{200}$ obtained from fitting the shear profile and its correspondence M$_{200}$. 
\end{flushleft}
\end{table*}

\subsection{A1204}

For this system we obtain the lensing masses of both identified substructures and the ratio between derived masses is $\sim 2$. \citet{Babyk2012} computed the mass for A1204 by using X-ray CHANDRA observations and assuming a NFW profile, obtaining $M_{200} = 3.18^{+0.34}_{-0.24} \times 10^{14} h^{-1}_{70} M_\odot $. This result is in good agreement with the estimated mass in this work ($M_{200} =  4.0 \pm 1.8 \times 10^{14} h^{-1}_{70} M_\odot$).\\
In the 2D density distribution (Fig. \ref{fig:mass-map}, \textit{upper panel}) we can distinguish only the primary structure. Taking into account the measured mass according to the fitted shear profile for this component, we compute the corresponding projected density profile, $\Sigma(r)$, assuming a NFW distribution according to the equations given by \citet{Wright2000}. All of the obtained values are below the 3$\sigma$ detection level (250 h$_{70} M_\odot/pc^2$) and only the inner region, corresponding to a radius of 60\,kpc would present a density larger than the 2$\sigma$ detection level (180 h$_{70} M_\odot/pc^2$). Thus, we consider that the mass of this structure is near the detection threshold of the 2D density contrast distribution. Also, we do not detect significant X-ray emission in the secondary structure region above the threshold adopted to build the brightness contours. In order to establish if this is due to the observing detection limit, we consider the lowest detected flux of RASS-based catalogs according to \citet{Piffaretti2011}, which contains the lowest X-ray emission clusters identified using this data. This corresponds to a flux F$_{lim}$ = $1.5 \times 10^{-12}$\,erg\,s$^{-1}$\,cm$^{-2}$, in the 0.1-2.4 keV band of the SGP catalog \citep{Cruddace2002}.  Considering this flux and the redshift of the secondary structure, we obtain a limiting luminosity of $1.5 \times 10^{42}$\,h$^{-2}_{70}$\,erg\,s$^{-1}$. Taking into account the obtained lensing mass, we compute the expected X-ray luminosity for the system according to \citet{Leauthaud2007} $M\_L_X$ relation, obtaining $(3.7 \pm 2.4) \times 10^{43}$\,h$^{-2}_{70}$\,erg\,s$^{-1}$, only $1.5\sigma$ above the detection limit. For this calculation we use the derived SIS mass since it has the lowest $\chi^2$ value and errors correspond to the propagation of the $M\_L_X$ parameter errors. Therefore, according to the lensing estimated mass the secondary structure would be roughly at the detection limit of the X-ray observation. We considered using more sensitive X-ray data, therefore we checked in the available databases of CHANDRA and XMM; however this component is not present in the field-of-view of these surveys. Considering  the  X-ray luminosity emission upper limit ($6.1 \times 10^{43}$\,h$^{-2}_{70}$\,erg\,s$^{-1}$) the emission is $2.5\sigma$ above the detection limit given by F$_{lim}$ \citep{Piffaretti2011}. Therefore, it is important to highlight that a low X-ray emissivity could be explained by a low density of the intracluster gas, which might be produced by a past interaction between the structures. \\
The cluster A1204 has recently been classified as a system with a strong cool core \citep{Zhang2016} which is in good correspondence with the X-ray contours  (Fig. \ref{fig:mass-map}, \textit{upper panel}). Thus, the cluster does not show any evidence of having suffered a recent merger event. Also, for this cluster the classification as a merging system is unstable as was stated in Section \ref{sec:dynamic}. Nevertheless, we detect lensing signal for the secondary structure through the lensing profile (Fig. \ref{fig:shear_profile_A1204}) from which we obtain the total mass, and, also, the density contours exhibited in Figure \ref{fig:mass-map} are elongated in the direction of the secondary component. Furthermore, assuming a relative velocity for these systems similar to the velocities measured in other merging clusters \citep[1000 km\,s$^{-1}$ ][]{Thompson2012}, the time necessary for the structures to reach the observed distance between the centers of the identified  substructures ($\sim 2.4$\,Mpc) is $\leq 3$\,Gyr.  Hence, taking into account that there are no other signatures of a collision, it is not likely that this could be a past merger event. Nevertheless this scenario could not be discarded if we were to consider lower velocities for the components; we  also do not discard the possibility that the interaction between these structures could be in process.
\subsection{A2029/A2033}
For this system we obtain the total mass for both structures which can be identified in the 2D projected density distribution. Both clusters show X-ray emission in good correspondence with the density distribution (Fig. \ref{fig:mass-map}). The obtained total mass for A2029 ($M_{200} =  9 \pm 6 \times 10^{14} h^{-1}_{70} M_\odot$) is in good agreement with that estimated by \citet{walker2012} using X-ray observations ($M_{200} = 10.1 \pm 0.6 \times 10^{14} h^{-1}_{70} M_\odot$). \\
A2029 is a relaxed galaxy cluster which has been studied extensively in X-rays  \citep[e.g., ][]{Sarazin1998, Lewis2002, Clarke2004, Vikhlinin2005, Snowden2008} . It  has  a  large cD  galaxy  \citep{Uson1991}  whose  major  axis  is aligned  in the NE to SW direction, in approximately the same direction as that joining it to nearby A2033. These two systems, together with A2028 and A2066 form a small supercluster \citep{Einasto2001}. Studies examining whether or not these systems are connected by a filament structure show that this is not the case \citep{walker2012,Planck2013}. Therefore, there is no conclusive evidence showing that these systems have interacted in the past or that they are now interacting, given the observed relaxed state of both structures. Nevertheless, if we take into account that they belong to the same supercluster and that the dynamical classification as interacting system is stable (see Sect. \ref{sec:dynamic}), it could be expected that these clusters would interact.

\subsection{Two-body model}

\begin{table*}

\caption{Solutions for the two-body model.}
\label{table:model}

\begin{tabular}{@{}ccrrrrcc@{}}
\hline
\rule{0pt}{1.05em}%
   Galaxy system  & Solution & $\alpha$  & $R$ & $R_m$ & $V$ & $P$   \\
                 &           & [deg]  & [$h^{-1}_{70} $\,Mpc] & [$h^{-1}_{70} $\,Mpc] & [km\,s$^{-1}$] & $\%$   \\ \hline
\rule{0pt}{1.05em}%
A1204  & Bound-outgoing         &   69.8   & 6.9  & 7.0 & 70 & 10 \\ 
       & Bound-incoming$_a$     &   66.9   & 6.1  & 6.1 & 70 & 19 \\ 
       & Bound-incoming$_b$     &    3.6   & 2.4  & 4.4 & 1000 & 71 \\ 
A2029/2033  & Bound-outgoing    &   75.7   & 13.6 & 64.6 & 690 & 100.  \\ 
\hline         
\end{tabular}
\medskip
\begin{flushleft}
\textbf{Notes.} Columns: (1) Cluster identification. (2) Solution class. Resultant values of $\alpha$ (3), $R$ (4), $R_m$ (5), $V$ (6) and the computed probability $P$ (7) for each solution. 
\end{flushleft}
\end{table*}

\begin{figure*}
\centering
\includegraphics[scale=0.45]{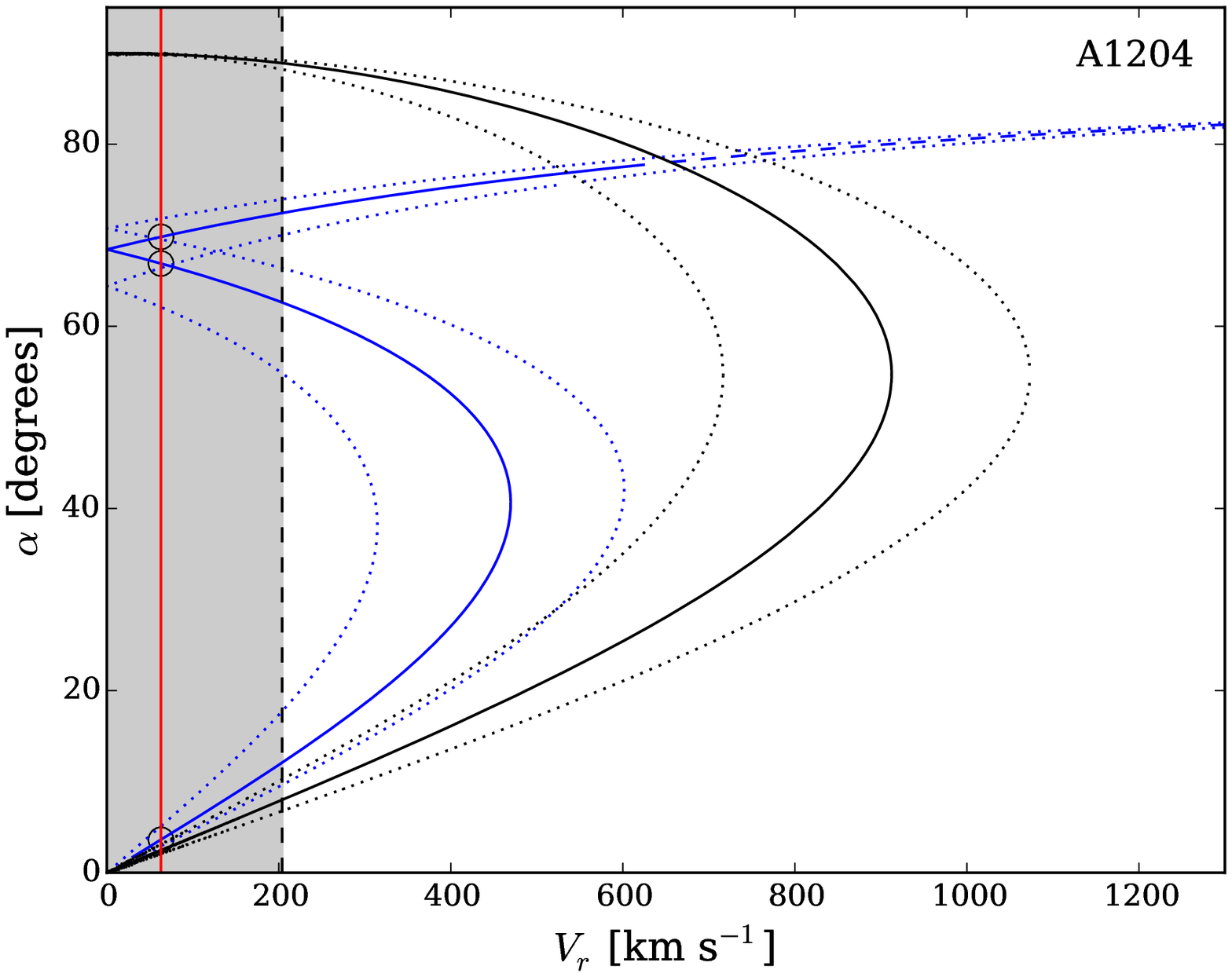}
\includegraphics[scale=0.45]{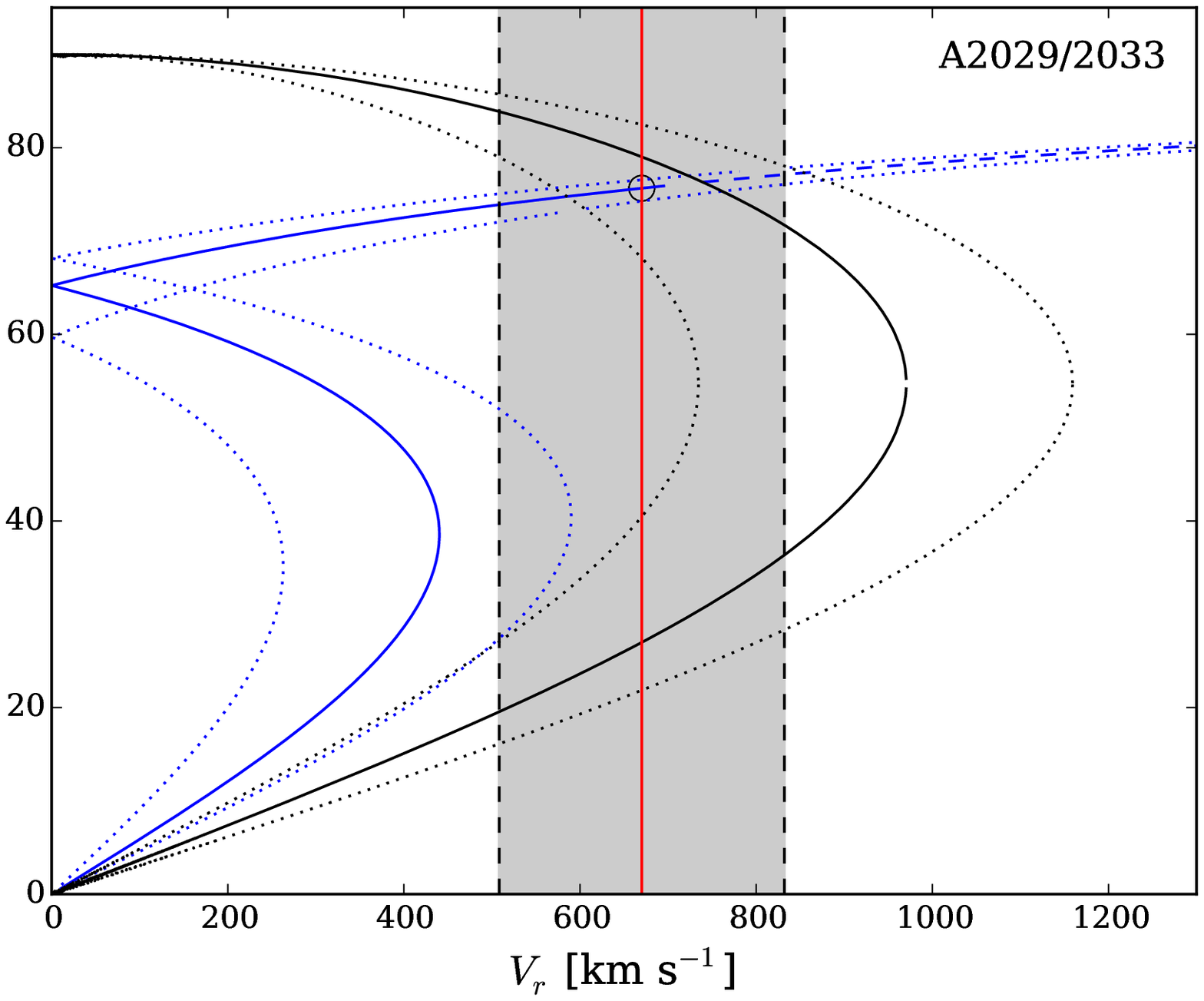}

      \caption{Projection angle ($\alpha$) as a function of the relative radial velocity difference ($V_r$) computed according to the equations of motion for A1204 (\textit{left pannel}) and A2029/2033 (\textit{right pannel}). Solid black curves separate the bound and unbound solutions according to the Newtonian criterion. Blue curves are the solutions of the equations of motions for bound (solid line) and unbound systems (dashed line) and the solutions for each system are marked with open circles, according to the observed $V_r$ marked with the red line where the gray region corresponds to the uncertainty in these values delimited by the dashed black line. Point curves express the uncertainties in the computed curves considering the errors in the adopted lensing masses.
}
\label{fig:Vr_alpha}
\end{figure*}

In order to obtain information regarding the state of evolution of the studied systems, we apply a Newtonian gravitational binding criterion that the two-body system is bound if the potential energy of the system is equal to or greater than the kinetic energy. This dynamical model was described in detail by \citet{Beers1982} and \citet{Gregory1984} and also applied to the analysis of several bimodal galaxy systems \citep[eg., ][]{Cortese2004,Hwang2009,Yan2014,Andrade20152,Caglar2017}. This model assumes radial orbits for the identified structures, which start their evolution at time $t_0 = 0$ with separation $R_0 = 0$, and are moving apart or coming together for the first time in their history. With the obtained lensing masses, this method allows us to estimate the probability that (1) the system is bound but still expanding, (2) the system is collapsing, or (3) the two structures are not bound to each other but are merely close together on the sky. It is important to highlight that this model does not consider the angular moment of the system, the distribution of matter inside each cluster, or the gravitational interaction of the infalling matter outside the cluster pair, since it is assumed that the masses are constant since their formation time \citep{Nascimento2016}. Nevertheless, this model puts another constraint on the dynamical state and the evolution of the systems.\\
The Newtonian criterion for gravitational binding can be stated in terms of the projected separation, $R_p$, the radial velocity difference, $V_r$ and the total mass of the combined system, $M$, as:
\begin{equation} \label{eq:newton}
V_r^2 R_p \leq 2GM sin^2( \alpha) cos( \alpha)
,\end{equation}
where $\alpha$ is the projection angle between the plane of the sky and the line connecting the components of the system. We can express the true (3D) velocity, $V$, and separation, $R$,  as:
\begin{equation}
R_p = R\, cos( \alpha), V_r = V\, sin( \alpha)
.\end{equation}
We compute $V_r$ considering the redshifts of the galaxy members of each component within the $R_{200}$ radius, computed according to SIS masses. $R_p$ is computed as the distance between each BCG (which is the adopted center for the lensing analysis) and the combined mass of the system is obtained by adding the SIS masses of each structure, since these have the lower errors. 

Using \citet{Beers1982} equations of motion for unbound and bound systems, we can obtain the projected angle $\alpha$ as a function of the radial velocity difference $V_r$. The equation relates the time (which is assumed for each system as the age of the Universe at the mean redshift of the considered structures), the velocity, $V$, and the separation, $R$, with the developmental angle, $\chi$, and the maximum separation, $R_m$, of the system components for the bound solution and the asymptotic expansion velocity for the unbound solution. The obtained $V_r - \alpha$ relations are plotted in Figure \ref{fig:Vr_alpha}. The solid black curve separates the bound and unbound solutions according to the Newtonian criterion (Eq. \ref{eq:newton}), the blue curves are the solutions of the equations of motions for bound (solid line) and unbound systems (dashed line) and the solutions for each system are marked with open circles, according to the observed $V_r$. For both systems the solutions are bound, defined by the intersections between the solutions of the equations of motion and the $V_r$. For each solution, $i$, we compute the probabilities given by:
\begin{equation}
p_i = \int^{\alpha_{sup,i}}_{\alpha_{inf,i}} cos (\alpha) d\alpha
,\end{equation}
then we normalize the obtained probability to obtain $P_i= p_i/(\sum_i p_i)$. \\
Solutions are presented in Table \ref{table:model}. For A2029 we obtain one bound-outgoing solution, close to the unbound solution considering the errors in $V_r$. According to this, the components are expanding separated by  a distance of $13.6\, h^{-1}_{70} $\,Mpc. On the other hand, for A1204 we obtain three bound solutions, two incoming and one outgoing. The most probable is the incoming solution for which the components have reached their maximum distance of $4.4 \,h^{-1}_{70} $\,Mpc and are now separated by $2.4\, h^{-1}_{70} $\,Mpc collapsing with a velocity of $1000$\,km\,s$^{-1}$.

\section{Summary and conclusions}
\label{sec:conc}
In this work we present the analysis of two candidates for interacting galaxy systems, A1204 and A2029/2033. For this analysis we include spectroscopic information of the galaxies identified as members, X-ray observations, and a weak-lensing study. \\
The dynamical analysis suggests that A1204 classification as a merging system is not stable and that each detected substructure is dynamically identified as not being in the process of merging. From the lensing analysis we were able to obtain the total mass of each component and the 2D projected density map. In the latter we observe a concentric distribution of the mass centered at the BCG of A1204, compatible with the gas distribution, according to the X-ray brightness contours. Therefore, this component is relaxed, hence there is no evidence that A1204 has passed through a recent merger event. Nevertheless, according to the two-body model, the identified structures are bound and collapsing. Therefore the identified components could interact in the future. \\
On the other hand, A2029/2033 classification as a merging system is stable according to the dynamical analysis and each cluster is classified separately as not being in the process of merging. By using our lensing analysis we derived the masses for each cluster and both components are visible in the 2D projected density distribution. The gas distribution given by the X-ray contours are in good agreement with the mass distribution and both systems seem to be relaxed. Hence, we conclude that for this system there is no evidence that these clusters have passed through a recent merger event. Furthermore considering the result of the two-body model, these clusters are not in the process of merging, since the solution describes an outgoing motion for the components.\\
Although the analyzed systems were classified as candidates for interacting galaxy clusters, they do not show evidence of having passed through a recent merger event. However, A1204 seems to be in process of interaction.  Thus in order to accurately classify the dynamical state of the systems, more constraints need to be taken into account. With this purpose we will continue the analysis of the identified systems in order to improve the algorithms of detection and also to provide a better understanding of the classified systems. Nevertheless, we highlight the performance of the merging system identification algorithm and the forthcoming works that would be based on its resultant catalog, since it would provide a uniform selected and analyzed sample of these interesting galaxy systems.

\begin{acknowledgements}
We thank the anonymous referee for the useful comments and suggestions which helped to improve the manuscript.\\
This research is based on data collected at Subaru Telescope, which is operated by the National Astronomical Observatory of Japan and on observations obtained with MegaPrime/MegaCam, a joint project of CFHT and CEA/IRFU, at the Canada-France-Hawaii Telescope (CFHT) which is operated by the National Research Council (NRC) of Canada, the Institut National des Science de l'Univers of the Centre National de la Recherche Scientifique (CNRS) of France, and the University of Hawaii. This work is based in part on data products produced at Terapix available at the Canadian Astronomy Data Centre as part of the Canada-France-Hawaii Telescope Legacy Survey, a collaborative project of NRC and CNRS.\\
Ww have made use of data and/or software provided by the High Energy Astrophysics Science Archive Research Center (HEASARC), which is a service of the Astrophysics Science Division at NASA/GSFC and the High Energy Astrophysics Division of the Smithsonian Astrophysical Observatory. We have made use of the ROSAT Data Archive of the Max-Planck-Institut für extraterrestrische Physik (MPE) at Garching, Germany.\\
We have also use SDSS-III data. Funding for SDSS-III has been provided by the Alfred P. Sloan Foundation, the Participating Institutions, the National Science Foundation, and the U.S. Department of Energy Office of Science. The SDSS-III web site is http://www.sdss3.org/.
SDSS-III is managed by the Astrophysical Research Consortium for the Participating Institutions of the SDSS-III Collaboration including the University of Arizona, the Brazilian Participation Group, Brookhaven National Laboratory, Carnegie Mellon University, University of Florida, the French Participation Group, the German Participation Group, Harvard University, the Instituto de Astrof\'isica de Canarias, the Michigan State/Notre Dame/JINA Participation Group, Johns Hopkins University, Lawrence Berkeley National Laboratory, Max Planck Institute for Astrophysics, Max Planck Institute for Extraterrestrial Physics, New Mexico State University, New York University, Ohio State University, Pennsylvania State University, University of Portsmouth, Princeton University, the Spanish Participation Group, University of Tokyo, University of Utah, Vanderbilt University, University of Virginia, University of Washington, and Yale University. \\
We made an extensive use of the following python libraries: http://www.numpy.org/, http://www.scipy.org/, http://roban.github.
com/CosmoloPy/ and http://www.matplotlib.org/.\\
This work was partially supported by the Consejo Nacional de Investigaciones Cient\'ificas y T\'ecnicas (CONICET,
Argentina) and the Secretar\'ia de Ciencia y Tecnolog\'ia de la Universidad Nacional de C\'ordoba (SeCyT-UNC, Argentina).\\
JLNC is grateful of the Southern Office of Aerospace Research and Development (SOARD), a branch of the Air Force Office of the Scientific Researchs International Office of the United States (AFOSR/IO), for the financial support obtained through grant FA9550-15-1-0167. JLNC also acknowledges financial support from the Direcci\'on de Investigaci\'on de la Universidad de La Serena through the Programa de Incentivo a la Investigaci\'on Acad\'emica (PIA-DIULS).

\end{acknowledgements}

\bibliography{references}

\end{document}